\documentclass[aip,jcp,amsmath,amssymb,reprint]{revtex4-1}
\usepackage{dcolumn}
\usepackage{bm}
\usepackage[latin1]{inputenc}
\usepackage{amsmath}
\usepackage{amsfonts}
\usepackage{amssymb}
\usepackage{bbold}
\usepackage{graphicx}
\usepackage{subfigure}

\begin{document} 
\preprint{AIP/123-QED} 
\title[Effects of noise-induced coherence on the fluctuations of current in quantum absorption refrigerators]
{Effects of noise-induced coherence on the fluctuations of current in quantum absorption refrigerators} 
\author{Viktor Holubec}
\email{viktor.holubec@mff.cuni.cz}
\affiliation{ 
Institut f{\"u}r Theoretische Physik, 
Universit{\"a}t Leipzig, 
Postfach 100 920, D-04009 Leipzig, Germany
}
\affiliation{Charles University, Faculty of Mathematics and
  Physics, Department of Macromolecular Physics, V Hole{\v s}ovi{\v
    c}k{\' a}ch 2, CZ-180~00~Praha, Czech Republic}

\author{Tom{\' a}{\v s} Novotn{\' y}} 
\email{tno@karlov.mff.cuni.cz.cz} 
\affiliation{Charles University, Faculty of Mathematics and
  Physics, Department of Condensed Matter Physics, Ke Karlovu 5, CZ-121 16 Praha, Czech Republic}
\date{\today} 
\begin{abstract} 
We investigate the effects of noise-induced coherence on average current and current fluctuations in a simple model of quantum absorption refrigerator with degenerate energy levels. We describe and explain the differences and similarities between the system behavior when it operates in the classical regime, where the populations and coherences in the corresponding quantum optical master equation decouple in a suitably chosen basis, and in the quantum regime, where such a transformation does not exist. The differences between the quantum and the classical case are observable only close to the maximum current regime, where the system steady-state becomes non-unique. This allows us to approximate the system dynamics by an analytical model based on a dichotomous process, that explains the behavior of the average current both in the classical and in the quantum case. Due to the non-uniqueness, the scaled cumulant generating function for the current at the vicinity of the critical point exhibits behavior reminiscent of the dynamical first-order phase transition. Unless the system parameters are fine-tuned to a single point in the parameter space, the corresponding current fluctuations are moderate in the quantum case and large in the classical case. 
\end{abstract} 
\maketitle  
\section{Introduction}  

One of the main motivations of quantum thermodynamics
is to determine whether quantum effects can be used to build machines operating better than the classical ones \cite{Dorfman2013, DelCampo2014, Vinjanampathy2016, Ghosh2019, Binder2019}. A machine operates in a truly quantum regime if its dynamics inseparably contains effects of quantum interference~\cite{Ghosh2019}. One of the simplest setups fulfilling this condition are open quantum systems with degenerate energy levels communicating with several heat reservoirs at different temperatures. Departing from an arbitrary initial condition, the non-equilibrium stationary states attained by these systems at long times contain non-vanishing coherences that cannot be in general accounted for by a special choice of the basis~\cite{Holubec2018, Gonzalez2019}. Because the steady-state coherences appear only if the temperatures, or noise intensities, differ, this effect has been called as the noise-induced coherence \cite{Scully2011,Svidzinsky2011,Svidzinsky2012,Dorfman2013}. 

The mechanism of noise-induced coherence is based on Fano interference \cite{Fano1961} corresponding to the transitions to and from the degenerate energy levels. For certain parameters, this interference leads to a division of the steady state into a dark state or blocking state \cite{Muralidharan2007}, that is separated from the rest of the system, and another steady state. This effect underlies various surprising quantum effects like lasing without inversion \cite{Harris1989} or electromagnetically induced transparency \cite{Boller1991,Rybin2015} and, surprisingly, can be to a large extent realized also using classical systems \cite{Joe2006}.

Investigations of the effect of the noise-induced coherence on the performance of quantum heat engines have been pioneered by Scully et al.~\cite{Scully2011,Svidzinsky2011,Svidzinsky2012,Dorfman2013}, who
reported a surprising enhancement of the power output in quantum heat engines with (nearly) degenerate energy levels. Subsequent works \cite{Creatore2013,Chen2016} addressed the status of the used quantum optical master equation and found similar effects using a classical Pauli rate equation. Nevertheless, our recent study~ \cite{Holubec2018} addressing averaged thermodynamic properties of quantum absorption refrigerators shows that systems exhibiting the noise-induced coherence can operate both in a classical regime, where the quantum description transforms into a classical one for a suitably chosen basis, and in a genuine quantum regime, where such a transformation is not possible.

The only advantage of the genuine quantum regime over the classical one found in Ref. \cite{Holubec2018} was the size of the parameter regime for which the refrigerator operated with maximum average cooling flux. In this work, we present a complementary analysis by further explaining the differences between the quantum and classical regime of operation and by investigating fluctuations of the flux in the considered absorption refrigerator. Current fluctuations in quantum and classical systems with degenerate steady-states exhibit behavior reminiscent of first-order phase transition \cite{Manzano2018}. We mainly focus to answer the following two questions. (i) Can this non-standard behavior of current fluctuations be beneficial for the performance of the quantum absorption refrigerator? (ii) Is the phase transition occurring in the quantum regime qualitatively different from that in the classical regime?

Although the concept of absorption refrigerators was put forward already in 1858 by F. Carr{\'e}, it now experiences its renaissance in the quantum realm \cite{Correa2013,Brask2015,Correa2014,Silva2015,Gonzalez2017,Segal2018,Kilgour2018,Mitchison2019} mainly initiated by the work of Levy and Kosloff \cite{Levy2012}. The main reasons are twofold. First, unlike cyclically operating devices that must be driven~\cite{Holubec2018b}, the absorption refrigerators operate autonomously in a time-independent non-equilibrium steady state~\cite{Pietzonka2018}. Hence both their experimental realization and theoretical studying is simpler than in case of the cyclic devices making them a perfect system for studying laws of quantum thermodynamics. Second, with the advent of quantum computation, it is now important to study microscopic quantum devices that can be potentially implemented inside quantum computers.

In the next Sec.~\ref{sec:model}, we first review the model of the absorption refrigerator used in Ref.~\cite{Holubec2018}. Afterward, in Sec.~\ref{sec:SSCF}, we review the definitions of average thermodynamic currents in the refrigerator. In Sec.~\ref{sec:SS}, we describe the structure of the steady-state attained by the refrigerator at long times and explain the differences between the quantum and the classical regime of operation of the system. Sec.~\ref{sec:CS} contains the main new results of the present paper concerning the statistics of the current in the absorption refrigerator with degenerate energy levels. Finally, we conclude in Sec.~\ref{sec:Concl}.

\section{Model}  
\label{sec:model}

\begin{figure}
\includegraphics[width=0.7\columnwidth]{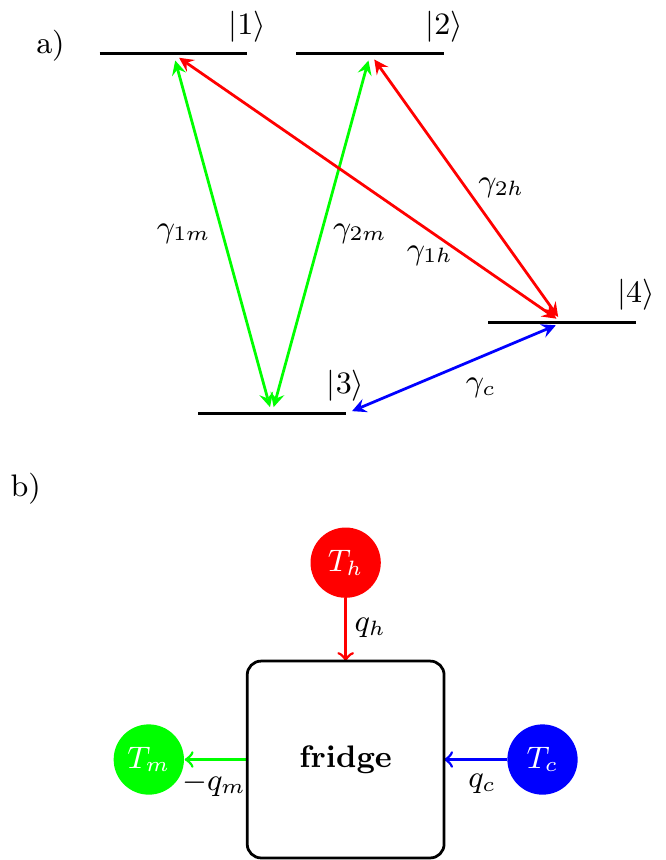}	
\caption{a) The level structure of the working medium of the considered absorption refrigerator schematically depicted in the panel b).}
\label{fig:fridges}
\end{figure}

We consider a quantum system with the structure of energy levels $\left|i\right>$ with energies $E_i = \hbar\omega_i$ depicted in Fig.~\ref{fig:fridges}~a). We assume that the energy levels $\left|1\right>$ and $\left|2\right>$ are degenerate and that the red, green and blue transitions in Fig.~\ref{fig:fridges}~a) are caused by coupling the system to heat reservoir at temperate $T_h$, $T_m$ and $T_c$, respectively. The temperatures $T_h\gg T_m>T_c$ and the energies $E_1 = E_2 > E_4 > E_3$ can be adjusted in such a way that the system utilizes the heat flow from the bath at $T_h$ to the one at $T_m$ to further cool the cold bath, as depicted in Fig.~\ref{fig:fridges} b). In this regime, the system works as an absorption refrigerator~\cite{Holubec2018} pumping heat both form hot and cold bath to the bath at the intermediate temperature $T_m$.

Assuming that the three reservoirs can be represented as black bodies described by a set of of harmonic oscillators coupled to the system by weak dipole interactions, the dynamics of the relevant elements $\rho_{ii} = \left<i\right|\rho\left|i\right>$, $i=1,\dots ,4$, and $\rho_R = [\left<1\right|\rho\left|2\right>  + \left<2\right|\rho\left|1\right>]/2$ of the system density matrix $\hat{\rho}(t)$ obey the following (approximate) quantum optical master equation
\begin{eqnarray}
\dot{\rho}_{11} &=&   
   \tilde{\gamma}_{1m} n_m^+ \rho_{33} + \tilde{\gamma}_{1h} n_h^+ \rho_{44}-(\tilde{\gamma}_{1h} n_h^- + \tilde{\gamma}_{1m} n_m^-)\rho_{11}  \nonumber\\  
 &-& (\tilde{\gamma}_{12h} n_h^- + \tilde{\gamma}_{12m} n_m^-)\rho_{R},
	\nonumber\\
\dot{\rho}_{22} &=&    \tilde{\gamma}_{2m} n_m^+ \rho_{33} + \tilde{\gamma}_{2h} n_h^+ \rho_{44} - (\tilde{\gamma}_{2h} n_h^- +
    \tilde{\gamma}_{2m} n_m^-)\rho_{22} \nonumber \\
   &-& (\tilde{\gamma}_{12h} n_h^- + \tilde{\gamma}_{12m} n_m^-) \rho_{R},
	 \nonumber\\
\dot{\rho}_{33} &=& \tilde{\gamma}_{c} n_c^- \rho_{44} + \tilde{\gamma}_{1m} n_m^- \rho_{11} + \tilde{\gamma}_{2m} n_m^- \rho_{22} \nonumber \\ &-&  (\tilde{\gamma}_{1m} n_m^+  + \tilde{\gamma}_{2m} n_m^+ + \tilde{\gamma}_c n_c^+) \rho_{33} +  
 2\tilde{\gamma}_{12m} n_m^- \rho_{R},
\nonumber\\
\dot{\rho}_{44} &=& \tilde{\gamma}_c n_c^+ \rho_{33} + \tilde{\gamma}_{1h} n_h^- \rho_{11} + \tilde{\gamma}_{2h} n_h^- \rho_{22} \label{eq:freq} \\
	&-&	(\tilde{\gamma}_{1h} n_h^+ + \tilde{\gamma}_{2h} n_h^+ + \tilde{\gamma}_c n_c^-) \rho_{44} + 2\tilde{\gamma}_{12h} n_h^- \rho_{R} ,
\nonumber\\
\dot{\rho}_R &=& \frac{\tilde{\gamma}_{12m}}{2}[
2 n_m^+ \rho_{33} - n_m^- (\rho_{11} + \rho_{22})
]
\nonumber \\ &+&
\frac{\tilde{\gamma}_{12h}}{2}[
2 n_h^+ \rho_{44} - n_h^- (\rho_{11} + \rho_{22})
]
\nonumber \\ &-&
\frac{1}{2} [(\tilde{\gamma}_{1h} + \tilde{\gamma}_{2h}) n_h^- + (\tilde{\gamma}_{1m} + \tilde{\gamma}_{2m}) n_m^-]\rho_{R},
\nonumber
\end{eqnarray}
that can be derived from the Redfield master equation~\cite{Breuer2002}. The only difference from the derivation of the standard quantum optical master equation, where populations $\rho_{ii}$ and coherences $\rho_R$ decouple, is that the coherences and populations become coupled via the terms proportional to the coefficients $\tilde{\gamma}_{12h}$ and $\tilde{\gamma}_{12m}$. This is because the rotating wave approximation, that leads to the decoupling of coherences in the standard case of a non-degenerate spectrum, does not decouple coherences and populations corresponding to the degenerate energy levels $\left|1\right>$ and $\left|2\right>$. The equation can be written in the Lindblad form and thus it preserves positivity and normalization of the density matrix. 

In Eq.~\eqref{eq:freq}, $n_x^- \equiv 1 + n_x^+$ and $n_x^+ \equiv n_x =  1/[\exp(\hbar \omega_x/k_{\rm B} T_x) - 1]$ denotes the average number of photons with frequency $\omega_x$ in the reservoir at temperature $T_x$, $x=h,m,c$. Namely, $\omega_{h} \equiv \omega_1-\omega_4$ corresponds to $T_h$, $\omega_{m} \equiv \omega_1-\omega_3$ to $T_m$, and $\omega_c \equiv \omega_4-\omega_3$ to $T_c$. The parameters 
\begin{equation}
\tilde \gamma_x = \frac{\omega_x^3}{6 \pi \hbar \epsilon_0 c^3}\gamma_x
\end{equation}
are determined by scalar products $\gamma_{1h,m} = |\mathbf{g}_{1h,m}|^2$, $\gamma_{2h,m} = |\mathbf{g}_{2h,m}|^2$, $\gamma_{c} = |\mathbf{g}_{c}|^2$ and $\gamma_{12h,m} = \mathbf{g}_{1h,m}\cdot \mathbf{g}_{2h,m}^\star$ of the matrix elements $\mathbf{g}_{1h} = e \left<1\right|\hat{\mathbf{r}}\left|4\right>$, $\mathbf{g}_{2h} = e \left<2\right|\hat{\mathbf{r}}\left|4\right>$, $\mathbf{g}_{1m} = e \left<1\right|\hat{\mathbf{r}}\left|3\right>$, $\mathbf{g}_{2m} = e \left<2\right|\hat{\mathbf{r}}\left|3\right>$ and $\mathbf{g}_{c} = e \left<4\right|\hat{\mathbf{r}}\left|3\right>$ of the system electric dipole moment operator $\hat{\mathbf{r}}$. The symbols $e,\hbar,c,\epsilon_0$ and $k_{\rm B}$ above denote the absolute value of the elementary charge, the Planck constant, the speed of light, the vacuum permittivity, and the Boltzmann constant, respectively. All details of the considered setup, the full derivation of Eq.~\eqref{eq:freq}, and the description of the regimes of its validity can be found in Ref.~\cite{Holubec2018}.

The elements of the electric dipole moment operator are in general complex vectors and thus the cross-product $\gamma_{12h}$ can assume any value inside the complex circle $\sqrt{\gamma_{1h}\gamma_{2h}}\exp(i\phi_{h})$ with the phase factor $\phi_{h}$, and similarly for $\gamma_{12m}$. The system dynamics is affected only by the relative phase $\theta = |\phi_{h}-\phi_{m}|$. The probability current through the system interpolates between a maximum and a minimum values attained for $\theta=0$ and $\theta = \pi$, respectively, as discussed in Ref.~\cite{Sedlak2018} for a slightly different system, which, however, exhibits qualitatively the same dynamics with respect to $\theta$ as the system considered here. Here, we investigate the system behavior near the maximum current regime and thus we already assumed that $\theta = 0$ in Eq.~\eqref{eq:freq}. Then the cross-products $\gamma_{12h}$ and $\gamma_{12m}$ are real, positive and bounded from above by $\sqrt{\gamma_{1h}\gamma_{2h}}$ and $\sqrt{\gamma_{1m}\gamma_{2m}}$, respectively.

For computational reasons, it is useful to rewrite the system~\eqref{eq:freq} in the from
\begin{equation}
\dot{\mathbf{\rho}} = \mathcal{R} \mathbf{\rho}
\label{eq:ME}
\end{equation}
for the vector $\mathbf{\rho} = [\rho_{11},\rho_{22},\rho_{33},\rho_{44},\rho_{R}]$. The elements $\mathcal{R}_{ij}$, $i,j = 1,\dots,5$, of the matrix $\mathcal{R}$ can be read out of the equation~\eqref{eq:freq}. Since the vector $\mathbf{\rho}$ contains not only populations (probabilities) $\rho_{ii}$ but also coherence $\rho_R$, the matrix $\mathcal{R}$ is not a standard rate matrix. In particular, the preservation of the normalization $\sum_{i=1}^4 \rho_{ii} = 1$ does not imply that $\sum_{i=1}^5 \mathcal{R}_{ij} = 0$ for all $j$, but instead $\sum_{i=1}^4 \mathcal{R}_{ij} = 1$. In Sec.~\ref{sec:SS}, we will see that in the parameter regime $\gamma_{1h} = \gamma_{2h} = \gamma_{12h}$ and $\gamma_{1m} = \gamma_{2m} = \gamma_{12m}$ the spectrum of the matrix $\mathcal{R}$ becomes degenerate and thus, in this special case, the considered system becomes non-ergodic.

\section{Steady state cooling flux}  
\label{sec:SSCF}

The system in Fig.~\ref{fig:fridges} operates as a refrigerator if the system goes from state $\left|3\right>$ to state $\left|4\right>$ more often than back. Differently speaking, the probability current
\begin{equation}
J_c(t) = \tilde{\gamma}_c [n_c^+ \rho_{33}(t) - n_c^- \rho_{44}(t)]
\label{eq:Jc}
\end{equation}
from state $\left|3\right>$ to state $\left|4\right>$, which determines the amount of heat $Q_c(t) = (E_4-E_3) J_c(t)$ taken on average per unit time from the cold bath and causing the $\left|3\right>\to \left|4\right>$ transition, must be positive.

In general, the current $J_c(t)$ depends on the initial state of the system and on time. It may also differ from the current 
\begin{multline}
J_h(t)=\tilde{\gamma}_{1h} [n_h^+ \rho_{44}(t) - n_h^-\rho_{11}(t)] +\\ \tilde{\gamma}_{2h} [n_h^+ \rho_{44}(t) - n_h^-\rho_{22}(t)] - \tilde{\gamma}_{12h} n_h^-\rho_{R}(t)
\label{eq:Jh}
\end{multline}
from $\left|4\right> \to \left|1\right>$ and $\left|2\right>$ caused by the reservoir at $T_h$ and
\begin{multline}
J_m(t)= \tilde{\gamma}_{1m} [n_m^+ \rho_{33}(t) - n_m^-\rho_{11}(t)] +\\ \tilde{\gamma}_{2m} [n_m^+ \rho_{33}(t) - n_m^-\rho_{22}(t)]  - \tilde{\gamma}_{12m} n_m^-\rho_{R}(t)
\label{eq:Jm}
\end{multline}
 from $\left|1\right>$ and $\left|2\right> \to \left|3\right>$ caused by the reservoir at $T_m$~\cite{Holubec2018}. However, at long times, the considered system attains a time-independent non-equilibrium steady state. In this steady state, the time derivative of the density matrix vanishes yielding the conservation law 
\begin{equation}
J = J_c=J_h=-J_m
\label{eq:Jss}
\end{equation}
for the time-independent stationary current.

From now on, we focus solely on this stationary state where all the quantities important for the thermodynamic description of the refrigerator are determined by the stationary current $J$. Namely, the average heat currents $Q_x$, from the reservoirs at temperatures $T_x$ are given by
\begin{eqnarray}
Q_c &=& (E_4-E_3)J,\\ 
Q_h &=& (E_1-E_4)J,\\ 
Q_m &=& (E_1-E_3)(-J),
\end{eqnarray}
and the average amount of entropy $\sigma = - \sum_x Q_x/T_x$ produced in the steady-state per unit time can thus be written as
\begin{equation}
\sigma = \left(- \frac{E_4-E_3}{T_c} -\frac{E_1-E_4}{T_h} + \frac{E_1-E_3}{T_m}\right) J.
\label{eq:sigma}
\end{equation}

\section{Structure of the steady state: Quantum vs. Classical}  
\label{sec:SS}

The exactly degenerate doublet in the level spectrum of the considered refrigerator induces couplings of the non-diagonal elements of the density matrix $\rho_R= \rho_{12}=\rho_{21}$ (coherences) to the diagonal ones $\rho_{ii}$ (populations) in Eqs.~\eqref{eq:freq} and \eqref{eq:ME}. If the reservoir temperatures differ, these couplings prevail also for steady-state density matrix attained by the system at long times. As a result, the steady state density matrix necessarily contains non-zero coherences $\rho_R$, which can be viewed as ``induced'' by the compound effect of noises with different intensities, hence the effect was called as noise-induced coherence~\cite{Scully2011,Svidzinsky2011,Svidzinsky2012,Dorfman2013}. 

\begin{figure}
\includegraphics[width=0.85\columnwidth]{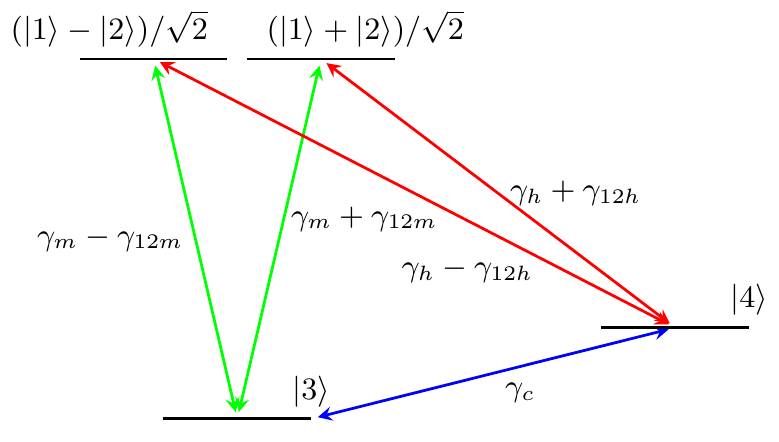}	
\caption{The level structure of the working medium of the considered absorption refrigerator in the transformed basis, where coherences decouple from populations. This transformation works only in the classical regime with $\gamma_{1x}=\gamma_{2x}$, $x=m,h$.}
\label{fig:system_fridge}
\end{figure}

Interestingly enough, in the \emph{symmetric} case $\gamma_{1h} = \gamma_{2h}$ and $\gamma_{1m} = \gamma_{2m}$, when the Fano interference for the transitions from the doublet is the strongest~\cite{Scully2011,Svidzinsky2011,Svidzinsky2012,Dorfman2013}, the master equation~\eqref{eq:freq} can be transformed into a classical form where the coherences and populations decouple, while such a transformation does not exist if the symmetry is broken~\cite{Holubec2018}. Below, we will refer to the symmetric situation as to the \emph{classical case} and to the setting with $\gamma_{1h} \neq \gamma_{2h}$ and/or $\gamma_{1m} \neq \gamma_{2m}$ as to the \emph{quantum case}.

Specifically, in the classical case, the coupling between the coherences and populations in the master equation~\eqref{eq:freq} vanishes after transforming the basis $\{\left|1\right>, \left|2\right>\}$ of the degenerate subspace to $\{(\left|1\right>-\left|2\right>)/\sqrt{2}, (\left|1\right>+\left|2\right>)/\sqrt{2}\}$. The system dynamics in the transformed basis is depicted in Fig.~\ref{fig:system_fridge}, where we have used the notation $\gamma_{1h} = \gamma_{2h} = \gamma_h$ and $\gamma_{1m} = \gamma_{2m} = \gamma_m$. The corresponding master equation $\dot{\rho'} = \mathcal{R}'\rho'$ for the transformed vector $\rho' = [(\rho_{11}+\rho_{22}-2\rho_R)/2,(\rho_{11}+\rho_{22}+2\rho_R)/2,\rho_{33},\rho_{44}, (\rho_{11}-\rho_{22})/2]$ can be obtained from Eq.~\eqref{eq:freq} by substituting $\gamma_{1m,h}$ by $\gamma_{m,h}-\gamma_{12m,h}$, $\gamma_{2m,h}$ by $\gamma_{m,h}+\gamma_{12m,h}$, and setting $\gamma_{12m,h} = 0$. The resulting equation still contains both populations and coherences, and thus it actually describes a quantum system, but now the dynamics of the populations, which in this case determine the probability current $J$ and thus the system performance, is completely independent of coherences. Moreover, the coherences in this case exponentially decay with time and thus they are zero in the steady state. This means that, although formally quantum, the system dynamics in the regime that we call as classical can be (almost) completely mimicked by a classical system described by the rate equation $\dot{p} = \mathcal{C} p$, where $p = [(\rho_{11}+\rho_{22}-2\rho_R)/2,(\rho_{11}+\rho_{22}+2\rho_R)/2,\rho_{33},\rho_{44}]$ and $\mathcal{C}_{ij} = \mathcal{R}'_{ij}$, $i,j=1,\dots,4$.
On  the other hand, if $\gamma_{1h} \neq \gamma_{2h}$ and/or $\gamma_{1m} \neq \gamma_{2m}$, no such a transformation that would allow us to separate coherences and populations in the master equation~\eqref{eq:freq} exists and thus it is impossible to mimic the corresponding dynamics by a classical rate equation. Such dynamics is hence genuinely quantum.

\begin{figure}[t!]
\includegraphics[width=0.95\columnwidth]{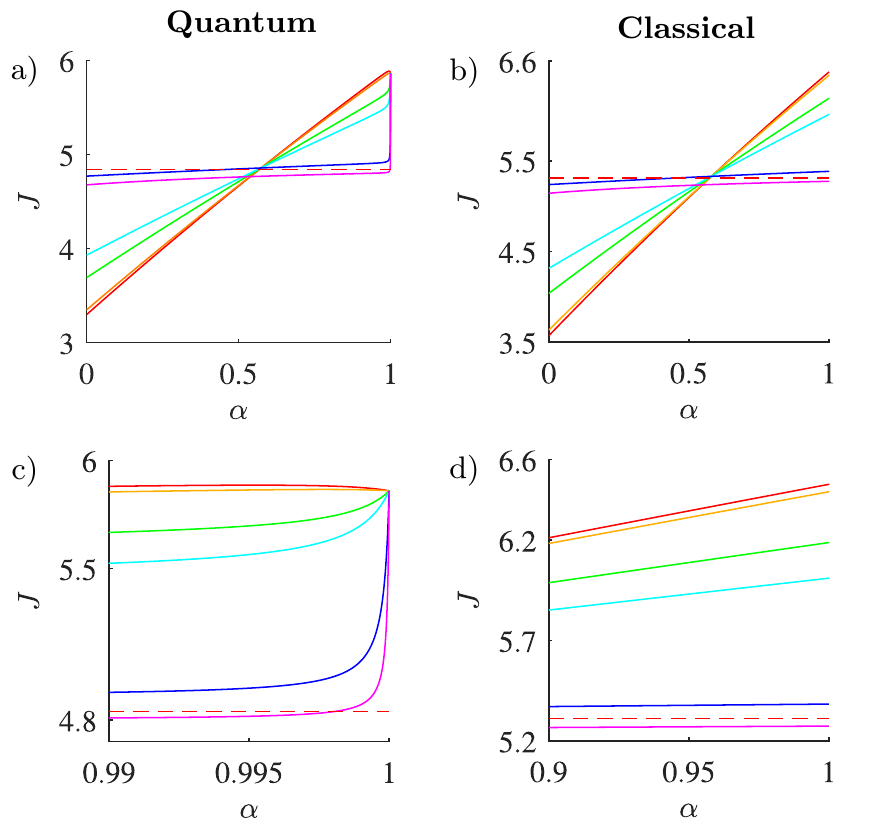}	
	\caption{The steady-state probability current $J$ for the six values 0, 0.001, 0.01, 0.02, 0.3 and 1 of the parameter $\sin\xi$ [full lines, from top to bottom in panel c)] as functions of $\alpha$ [see Eqs.~\eqref{eq:g12hfigs} and \eqref{eq:g12mfigs}]. The dashed lines were computed using $\gamma_{12h}=\gamma_{12m}=0$.  In panels a) and c), we show the current in the quantum regime with $\gamma_{1h} = 5.7 \gamma_0, \gamma_{2h} = 4.0 \gamma_0, \gamma_{1m} = 7.5 \gamma_0$, and $\gamma_{2m} = 6.0 \gamma_0$. The panels b) and d) correspond to the classical regime with $\gamma_{1h} = 5.7 \gamma_0, \gamma_{2h} = 5.7 \gamma_0, \gamma_{1m} = 7.5 \gamma_0$, and $\gamma_{2m} = 7.5 \gamma_0$. The parameter $\gamma_0 = (e\AA)^{2}\approx(5\, \mathrm{Debye})^{2}$ corresponds to a realistic value of the electric dipole element. We set $T_c = 34$ K, $T_m = 35$ K and $T_h = 5000$ K for the reservoir temperatures and $E_1 = E_2 = 0.01$ eV, $E_4 = 0.001$ eV and $E_3 = 0$ eV for the energies of the individual levels. We also always take $\gamma_c = 10^4 \gamma_0$.}	
	\label{fig:QMvsC}	
\end{figure}

The larger the cross-products $\gamma_{12x}$, $x=m,h$, the larger the steady state coherence $|\rho_R|$ and the current $J$. Hence, both $|\rho_R|$ and $J$   
reach their maximum values for $\gamma_{12x} \to \sqrt{\gamma_{1x}\gamma_{2x}}$ corresponding to the maximum Fano interference~\cite{Holubec2018}. Let us now investigate the regime close to the maximum current, where the system behaves slightly differently in the quantum and in the classical parameter regime, as reported in Ref.~\cite{Holubec2018}. To this end, we alter the parameters $\gamma_{12x}$ according to the formulas
\begin{eqnarray}
\gamma_{12h} &=& [1 - (1-\alpha)\sin\xi]\sqrt{\gamma_{1h}\gamma_{2h}}\ ,
\label{eq:g12hfigs}\\
\gamma_{12m} &=& [1 - (1-\alpha)\cos\xi]\sqrt{\gamma_{1m}\gamma_{2m}}\ ,
\label{eq:g12mfigs}
\end{eqnarray}
and compute the corresponding current. The results are shown in Fig.~\ref{fig:QMvsC}, where we plot the current $J$ as a function of the parameter $\alpha$
for six values of the parameter $\sin\xi$ corresponding to reaching the maximum coherence limit $\gamma_{12x} \to \sqrt{\gamma_{1x}\gamma_{2x}}$ (attained for $\alpha=1$) from several different directions in the $\gamma_{12m}$--$\gamma_{12h}$ plane. In the quantum regime, the limit $\lim_{\alpha \to 1} J$ does not depend on the direction (the individual full lines in panels a) and c) converge with $\alpha\to 1$ to the same value), while in the classical regime (panels b) and d)) the values of $\lim_{\alpha \to 1} J$ obtained for the individual parameters $\sin\xi$ differ. 

The non-uniqueness of the limit in the classical case can be traced to non-uniqueness (degeneracy) of the steady state in this regime~\cite{Holubec2018}, known in the field of quantum transport as a blocking state \cite{Muralidharan2007}. Namely, for $\gamma_{1x}=\gamma_{2x}$ and $\gamma_{12x}=\gamma_{1x}$, $x=m,h\ (\alpha=1)$, the null-space of the matrix $\mathcal{R}$ becomes two-dimensional. One of its components is the combination $(\rho_{11}+\rho_{22}-2 \rho_R)/2$ independent of the model parameters and disconnected from the rest of the system, thus sometimes called the dark state, which corresponds to the antisymmetric combination $|D\rangle\equiv (|1\rangle - |2\rangle)/\sqrt{2}$ in Fig.~\ref{fig:system_fridge}. We will now discuss the behavior of the classical system close to the degenerate case, i.e.\ when $\alpha\to 1$.  To illustrate the mechanism analytically, we adopt a limit of $n_{h}\to\infty$ and  $n_{m}\to 0$ which simplifies the expressions to manageable forms and still is within 10\% of the full numerical solution for our parameters (in our case $n_{h}\approx 50$ and $n_{m}\approx 0.04$). In this limit, the stationary state of the (nearly) decoupled three-state system involving states $\{|3\rangle,|4\rangle,(|1\rangle + |2\rangle)/\sqrt{2}\}$ is described by the vector $(1-2\pi_{0},\pi_{0},\pi_{0})$ with
\begin{equation}
\begin{split}
 \pi_{0}(\alpha)&=\frac{ \tilde{\gamma}_{c}n_{c}}{ \tilde{\gamma}_{c}(3n_{c}+1)+ \tilde{\gamma}_{m}[2-(1-\alpha)\cos\xi]}\\
                        &=\frac{\pi_{0}(1)}{1-\pi_{0}(1)\frac{ \tilde{\gamma}_{m}}{ \tilde{\gamma}_{c}n_{c}}(1-\alpha)\cos\xi} \\
                        &\approx \pi_{0}(1)\left[1+\pi_{0}(1)\frac{ \tilde{\gamma}_{m}}{ \tilde{\gamma}_{c}n_{c}}(1-\alpha)\cos\xi\right].   
\end{split}
\end{equation}
The associated current then reads 
\begin{align}
J_{0}(\alpha)&= \tilde{\gamma}_{m}[2-(1-\alpha)\cos\xi]\pi_{0}(\alpha)\\\nonumber
&\approx J_{\rm max}[1-s(1-\alpha)\cos\xi],
\intertext{with the maximum current}
J_{\rm max}&=2 \tilde{\gamma}_{m}\pi_{0}(1)=\frac{2 \tilde{\gamma}_{m} \tilde{\gamma}_{c}n_{c}}{ \tilde{\gamma}_{c}(3n_{c}+1)+2 \tilde{\gamma}_{m}}\label{eq:quantumJ}
\intertext{and the slope}
s &=\frac{1}{2}-\frac{\tilde{\gamma}_{m}\pi_{0}(1)}{ \tilde{\gamma}_{c}n_{c}}=\frac{1}{2}\frac{3n_{c}+1}{3n_{c}+1+\frac{2 \tilde{\gamma}_{m}}{ \tilde{\gamma}_{c}}}.
\end{align}

Next, we allow a weak (for $\alpha$ close to 1) connection between the dark state and the three-state cycle assuming a slow dichotomous switching between the dark state, blocking any current, and the three-state cycle, carrying current $J_{0}(\alpha)$ above. The switching process is governed by the switching rates
\begin{align}
\Gamma_{\uparrow} &=(1-\alpha)[ \tilde{\gamma}_{h}n_{h}^{+}\pi_{0}(\alpha)\sin\xi+\tilde{\gamma}_{m}n_{m}^{+}(1-2\pi_{0}(\alpha))\cos\xi]
\intertext{for the transition from the cycle to the dark state and}
\Gamma_{\downarrow} &=(1-\alpha)[ \tilde{\gamma}_{h}n_{h}^{-}\sin\xi+\tilde{\gamma}_{m}n_{m}^{-}\cos\xi]
\end{align}  
for the backward transition from the dark state. Most importantly, both rates are proportional to the distance from the critical point $\alpha=1$ and, consequently, in its vicinity they are very small setting the longest timescale in the problem. Then the proposed picture of the dichotomous switching between two dynamical states (the dark state with zero current and three-state cycle with a nonzero current $J_{0}(\alpha)$) is well justified not only for the average current but even for its long time fluctuations as we will demonstrate in the next section. Here, we just give the final expression for the total mean current after the averaging over the switching process. It reads
\begin{equation}\label{eq:stredni-proud}
\begin{split}
J(\alpha)&=\frac{\Gamma_{\downarrow}}{\Gamma_{\downarrow}+\Gamma_{\uparrow}}J_{0}(\alpha)\\
&\approx J_{0}(\alpha)\frac{\tilde{\gamma}_{h}(1+n_{h})\sin\xi+\tilde{\gamma}_{m}\cos\xi}{\tilde{\gamma}_{h}(1+n_{h}[1+\pi_{0}(\alpha)])\sin\xi+\tilde{\gamma}_{m}\cos\xi}\\
&\approx J_{\rm max}\frac{\tilde{\gamma}_{h}(1+n_{h})\sin\xi+\tilde{\gamma}_{m}\cos\xi}{\tilde{\gamma}_{h}(1+n_{h}[1+\pi_{0}(1)])\sin\xi+\tilde{\gamma}_{m}\cos\xi}\times \\ 
&\qquad[1-s'(1-\alpha)\cos\xi]\\
&\approx J_{\rm max}\frac{1+\frac{\tilde{\gamma}_{m}\cot\xi}{\tilde{\gamma}_{h}n_{h}}}{1+\pi_{0}(1)+\frac{\tilde{\gamma}_{m}\cot\xi}{\tilde{\gamma}_{h}n_{h}}}[1-s'(1-\alpha)\cos\xi],
\end{split}
\end{equation}
with the explicit expression for the final slope $s'>0$ 
\begin{equation}
\begin{split}
   s' &= s + \frac{\pi^{2}_{0}(1)\tilde{\gamma}_{m}}{\tilde{\gamma}_{c}n_{c}}\frac{1}{1+\pi_{0}(1)+\frac{\tilde{\gamma}_{m}\cot\xi}{\tilde{\gamma}_{h}n_{h}}} \\
     &=\frac{1}{2}\frac{4 n_c+1   +\left(3 n_c+1\right) \frac{\tilde{\gamma}_{m}\cot\xi}{\tilde{\gamma}_{h}n_{h}} }{4 n_c+1+2 \frac{\tilde{\gamma}_m}{\tilde{\gamma}_{c}}+ \left(3 n_c+1+2 \frac{\tilde{\gamma}_m}{\tilde{\gamma}_{c}}\right)\frac{\tilde{\gamma}_{m}\cot\xi}{\tilde{\gamma}_{h}n_{h}}}.
\end{split}
\end{equation}

Now, we can use the above-derived equation \eqref{eq:stredni-proud} for explaining the behavior observed in the right panel (classical results) of Fig.~\ref{fig:QMvsC}. We see that both the slope of the classical curves in the vicinity of at the critical point $\alpha=1$ as well as the values of the current there (Fig.~\ref{fig:QMvsC}d)) do depend on the value of the parameter $\xi$ consistently with the prediction of Eq.~\eqref{eq:stredni-proud}. In particular, the slope is (roughly, since $s'$ does also depend on $\xi$) proportional to $\cos\xi$ and the critical value interpolates between the maximum value of the current at $\sin\xi=0$ and $J_{\rm max}/(1+\pi_{0}(1))\approx 0.8J_{\max}$ at $\sin\xi=1$. Due to the large value of $n_{h}$, the crossover in $\xi$ is quite fast as is obvious from Fig.~\ref{fig:QMvsC}d).         

In the quantum case, the coupling of the dark state to the rest of the system, proportional to $\gamma_{1x}+\gamma_{2x}-2\gamma_{12x}$, $x=h,m$ \cite{Holubec2018}, does not reach zero even for $\alpha=1$. Moreover, there is an extra coupling between the two subsystems mediated by the remaining coherence (which does not fully decouple from the populations unlike in the classical case). The net result of these effects is that the dark state never fully separates from the rest of the system and, thus, the stationary state is always unique. Apparently, stemming from the numerical results in Fig.~\ref{fig:QMvsC} as well as those for fluctuations below, there is a possibility of an effectively classical dichotomous description with rates which, however, due to the quantum corrections saturate at non-zero values with $\alpha\to 1$. This preserves the uniqueness of the steady state as well as tames the fluctuations as we proceed to show in the next section. 

\section{Statistics of the current}  
\label{sec:CS}

We will focus on the statistics of the fluctuating current 
\begin{equation}
j = \frac{1}{t}\int_0^t dt' [n_{3\to 4}(t') - n_{4\to 3}(t')]
\label{eq:fluctj}
\end{equation}
defined as the difference between the number of transitions $n_{3\to 4}$ from the level $\left|3\right>$ to the level $\left|4\right>$ [$n_{3\to 4}(t')=\delta(t')$ if the transition occurred at time $t'$ and zero otherwise] and the number $n_{4\to 3}$ of backward transitions averaged over time $t$. For $t\to \infty$, the time averaged current equals to the mean current $J$. For large, but finite times $t$,  the statistics of the current $j$ can be described using the so-called large deviation theory~\cite{Touchette2009}.

\subsection{Large deviations of the current}  

Within the large deviation theory, one constructs the approximate probability distribution function (PDF)
\begin{equation}
\log P(t,j) \approx -t I\left(j\right)
\label{eq:rhoLDF}
\end{equation}
from the so-called large deviation function $I(j)$ that plays a role equivalent to the equilibrium free energy, and governs fluctuations of non-equilibrium currents \cite{Bertini2001,Touchette2009}. The symbol $\approx$ in the formula~\eqref{eq:rhoLDF} denotes that the two sides equal up to the corrections that are sublinear in time $t$. The large deviation function is usually obtained by the Legendre--Fenchel transform (see Refs.~\cite{Touchette2009,Holubec2018a} for details) 
\begin{equation}
I(j) = -\min_u [\lambda(u) + u j]
\label{eq:LDF}
\end{equation} 
from the so-called scaled cumulant generating function $\lambda(u)$, which
can be computed as the maximum eigenvalue of the so-called tilted matrix 
$\tilde{\mathcal{R}}(u)$. The tilted matrix $\tilde{\mathcal{R}}(u)$ is identical to the matrix $\mathcal{R}$ except for the elements responsible for changing the value of the current $j$. The element $\tilde{\gamma}_c n_c^+$, that is responsible for increase in the current $t j$ by 1, is multiplied by $\exp(-u)$ and, similarly, the element $\tilde{\gamma}_c n_c^-$, causing a unit decrease in the current, is multiplied by $\exp(u)$. These two exponential factors result as two-sided Laplace transforms of the probability distribution for the current $t j$ for a short-time evolution, see Ref.~\cite{Holubec2018a} for a more thorough discussion. Let us note that, for the present simple setting, the described intuitive method gives the same results as the method based on the dressed master equation used in Ref.~\cite{BulnesCuetara2016}.

\begin{figure}[t!]
\includegraphics[width=0.95\columnwidth]{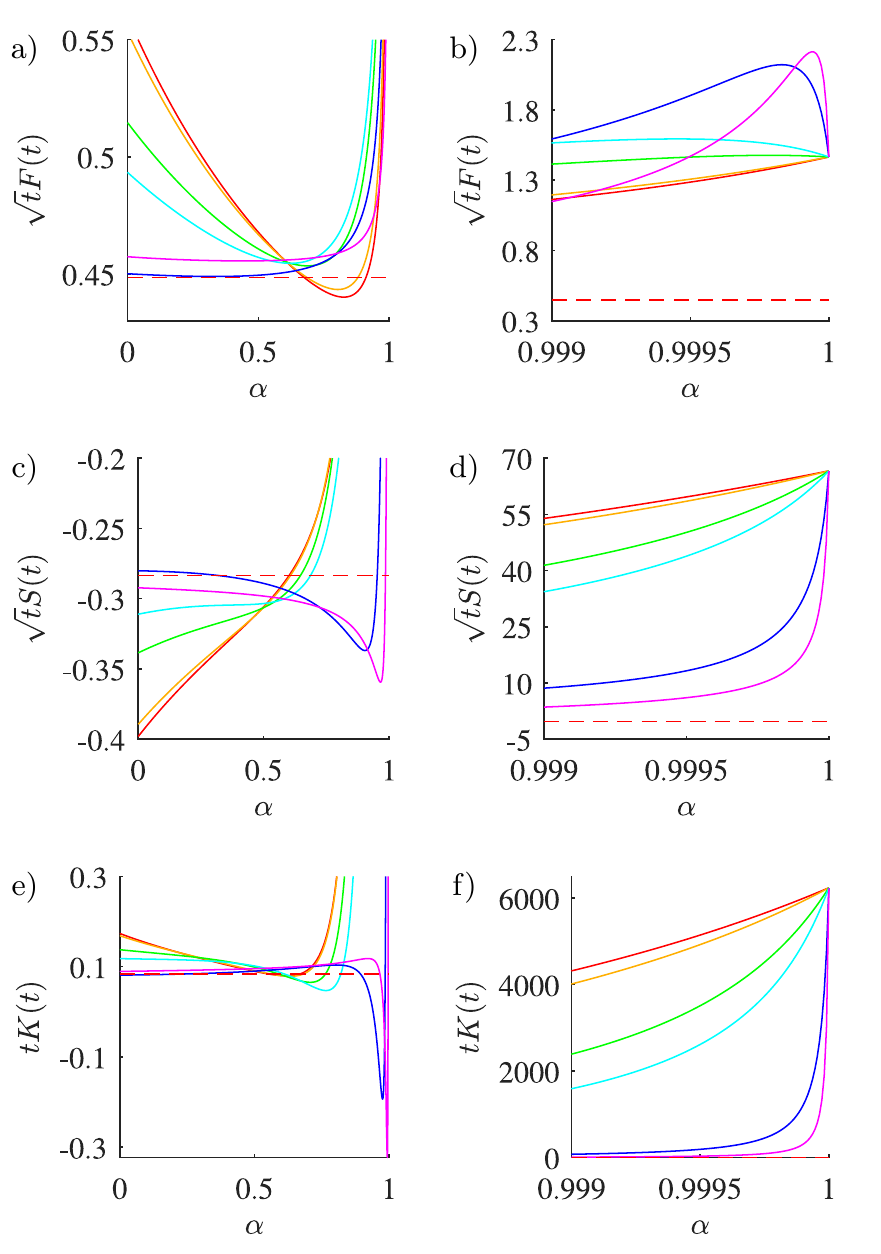}	
	\caption{Scaled relative current fluctuation~\eqref{eq:F} [a) and b)], skewness~\eqref{eq:S} [c) and d)] and kurtosis~\eqref{eq:K} [e) and f)] in the quantum regime. We used the same parameters as in Figs.~\ref{fig:QMvsC}~a) and c). The parameter $\sin\xi$ assumes the values 0, 0.001, 0.01, 0.02, 0.3 and 1 from top to bottom in panels d) and f).}	
	\label{fig:shape1}	
\end{figure}

\begin{figure}[t!]
\includegraphics[width=0.95\columnwidth]{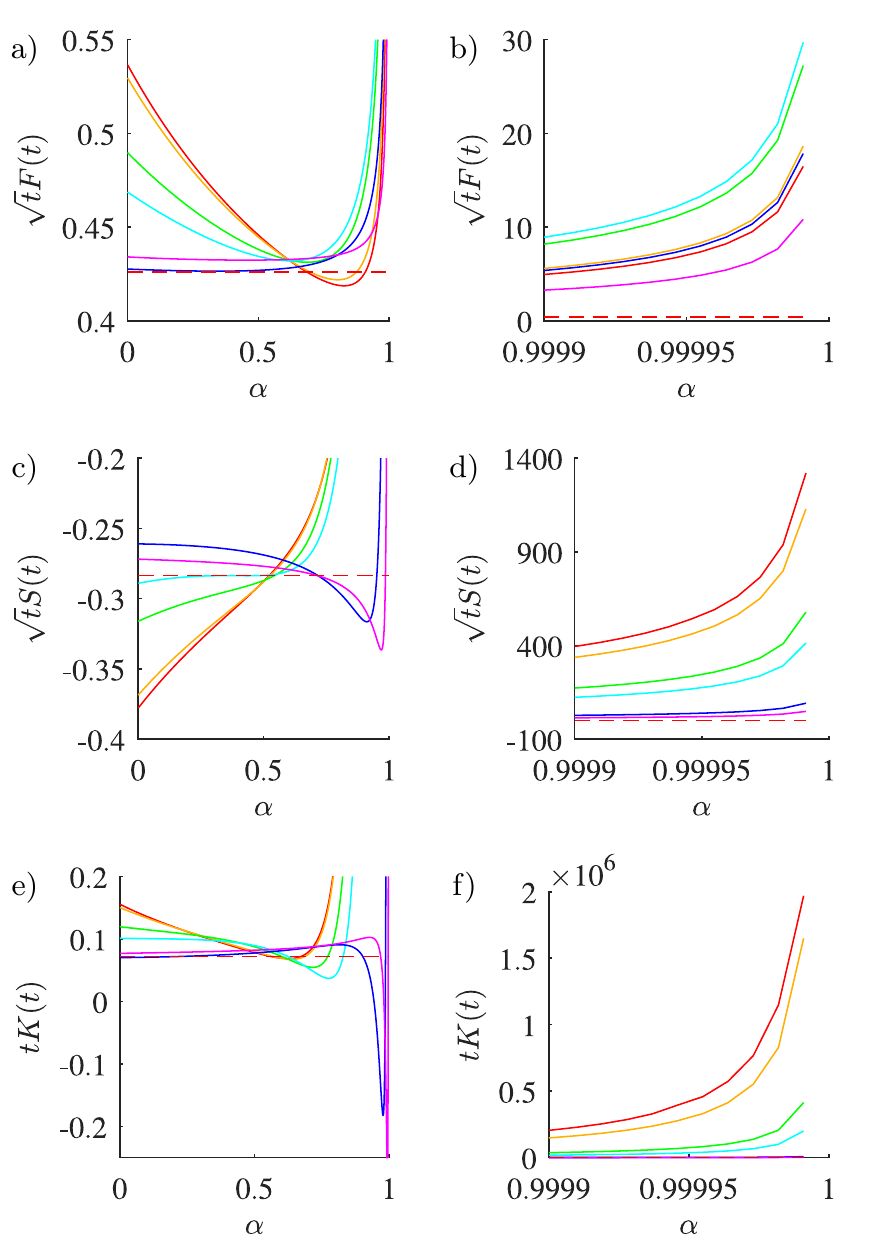}	
\caption{Scaled relative current fluctuation~\eqref{eq:F} [a) and b)], skewness~\eqref{eq:S} [c) and d)] and kurtosis~\eqref{eq:K} [e) and f)] in the classical regime. We used the same parameters as in Figs.~\ref{fig:QMvsC}~b) and d). The parameter $\sin\xi$ assumes the values 0, 0.001, 0.01, 0.02, 0.3 and 1 from top to bottom in panels d) and f).}	
	\label{fig:shape2}	
\end{figure}

To compare the behavior of the fluctuations in the quantum and classical regime described in the previous section, we use the scaled cumulant generating function to compute the scaled cumulants
\begin{equation}
\tilde{C}_k(t) = (-1)^k \frac{d^k \lambda(u)}{du^k}\bigg|_{u=0}
= t^{k-1} C_k.
\end{equation}
The factor $t^{k-1}$ relating the scaled cumulants to the cumulants $C_k$ of the current $j$ arises from the fact that $t\tilde{C}_k(t)$ is the $k$-th cumulant for the time-integrated current $t j$ (the pre-factor $t$ in $t \tilde{C}_k(t)$ comes from the definition of the scaled cumulant generating function~\cite{Holubec2018a}).

First, we have verified that the first cumulant $C_1 =  \tilde{C}_1 = J$ coincides with our results for $J$ obtained in the previous section. Afterward, we have evaluated the relative current fluctuation
\begin{equation}
F(t) = \frac{\sqrt{\left<j^2\right>-J^2}}{J} = \frac{\sqrt{C_2}}{C_1} = \frac{1}{\sqrt{t}} \frac{\sqrt{\tilde{C}_2}}{\tilde{C}_1},
\label{eq:F}
\end{equation}
the skewness
\begin{equation}
S(t) = \frac{C_3}{C_2^{3/2}} = \frac{1}{\sqrt{t}} \frac{\sqrt{\tilde{C}_3}}{\tilde{C}_2^{3/2}},
\label{eq:S}
\end{equation}
and the kurtosis
\begin{equation}
K(t) = \frac{\left<(j-J)^4\right>}{C_2^2} - 3 = \frac{C_4}{C^2_2} = \frac{1}{t}\frac{\tilde{C}_4}{\tilde{C}_2^2}.
\label{eq:K}
\end{equation}
Because all three above quantities decay to zero with time $t$, we show in Fig.~\ref{fig:shape1} (quantum case) and in Fig.~\ref{fig:shape2} (classical case) their time-rescaled variants which no longer depend on $t$. The individual lines in the figures correspond to the lines depicted in Fig.~\ref{fig:QMvsC}.

Besides the uniqueness of the $\alpha\to 1$ limit of the depicted coefficients in the quantum case and its non-uniqueness in the classical case, the main difference are the magnitudes of the moments in the individual situations. While the scaled relative fluctuation $F$ is in the classical case more than 10-times larger than in the quantum one, this factor increases to more than 23 for the scaled skewness and to 1000 for the scaled kurtosis. Hence, while the both current PDFs are relatively broad, positively skewed and with a positive kurtosis close to the critical point, these features are much more pronounced in the classical regime than in the quantum regime. Another difference can be spotted in panels b) of the two figures: close to the critical point in the quantum case, the relative fluctuation is smaller for larger values of the mean current. In the classical case, this hierarchy can be broken, as verified by the lowermost full (purple) line in Fig.~\ref{fig:shape2} b) that corresponds to the smallest value of the mean current in Fig.~\ref{fig:QMvsC} and largest relative fluctuation in Fig.~\ref{fig:shape1} b).
Both in the quantum and in the classical regime, the larger the average current, the larger the scaled skewness and scaled kurtosis close to the critical point.

\begin{figure}[t!]
\includegraphics[width=0.95\columnwidth]{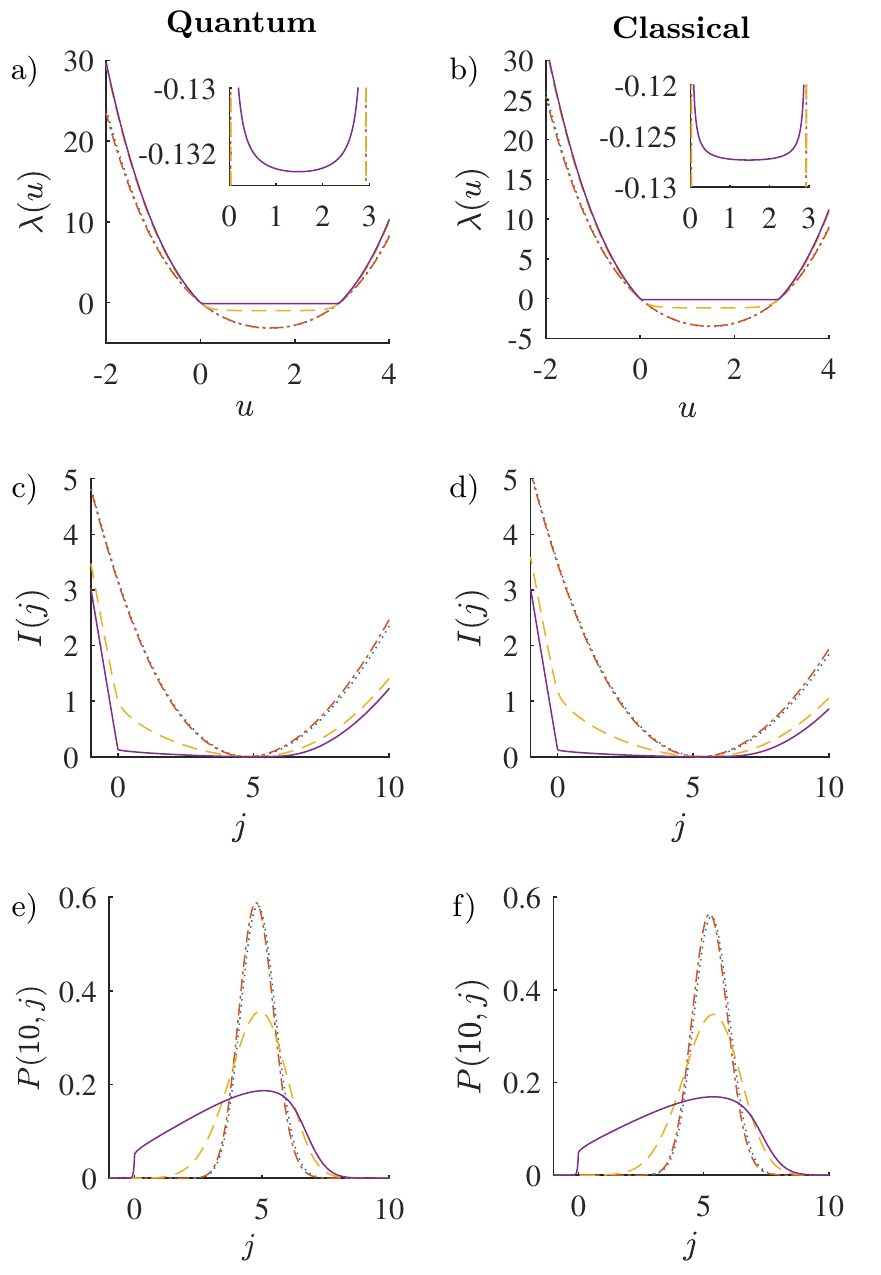}	
	\caption{Scaled cumulant generating functions [a) and b)], large deviation functions [c) and d)] and PDFs $P(t,j) = \exp[-t I(j)]/N$, $N=\int_{-\infty}^{\infty} dj \exp[-t I(j)]$, for $j$ at time $t=10$ [e) and f)]
obtained from the large deviation theory in the quantum regime (left) and in the classical regime (right). In the insets in the panels a) and b), we show in detail the seemingly flat regions of the cumulant generating functions. We took $\cos\xi=0.3$ and $\gamma_{12h}=\gamma_{12m}=0$ (dotted lines), $\alpha = 0$ (dot-dashed lines), $\alpha = 0.99$ (dashed lines) and $\alpha = 0.999$ (full lines). The dotted and the dot-dashed lines almost overlap. Other parameters for the quantum and the classical regime are the same as in Fig.~\ref{fig:QMvsC}.}	
	\label{fig:PDF}	
\end{figure}

Typical shapes of the scaled cumulant generating functions, large deviations functions, and the corresponding PDFs in the quantum and in the classical regime for four distances from the critical point are depicted in Fig.~\ref{fig:PDF}. While the depicted functions develop the same qualitative features, in the panels e) and f) for the quantum and the classical PDF, respectively, one can observe the above discussed larger relative fluctuation, skewness and kurtosis reached in the classical regime compared to the quantum one. 

\subsection{Dynamic phase transitions in current statistics}  

The most interesting parameter regime shown in the figures is the one closest to the critical point $\alpha=1$, depicted by the full lines. Both in the quantum and in the classical case, the corresponding scaled cumulant generating functions, shown in panels a) and b) of Fig.~\ref{fig:PDF}, develop almost constant region, connected to the rest of the curve by two points (at $u\approx 0$ and $u\approx 3$) where the derivative $\lambda'(u)$ changes very fast. This feature of $\lambda(u)$ then manifests itself through the Legendre--Fenchel transform in the steep increase of the large deviation functions [panels c) and d)], and thus also in the steep decrease of the PDFs [panels e) and f)]. The presence of the plateau in the scaled cumulant generating function and the corresponding steep decay of the PDFs (exactly) at $j=0$ is caused by the degeneracy of the steady state for $\alpha=1$ 
and $\gamma_{1x}=\gamma_{2x}$, $x=h,m$~\cite{Manzano2018}. In this limit, the plateau of the scaled cumulant generating function is completely flat and the derivative $\lambda'(u)$ develops a discontinuity at the borders of the plateau. The non-analytical scaled cumulant generating function is a hallmark of a first-order phase transition \cite{Garrahan2010}, where the Legendre--Fenchel transform \eqref{eq:LDF} can no longer be used for determination of the large deviation function and thus of the probability PDF~\cite{Touchette2009}. 

In our setup, the failure of the Legendre--Fenchel transform at the critical point can be easily understood. One of the assumptions behind the formula \eqref{eq:LDF} is that the large deviation function does not depend on the preparation of the system. While this assumption is fulfilled for parameters arbitrarily close to the critical point, i.e. for $\alpha<1$ and/or $\gamma_{1x}\neq \gamma_{2x}$, it breaks down at the critical point with $\alpha=1$ and $\gamma_{1x}= \gamma_{2x}$, where the steady-state becomes non-unique. Such a smooth transition of system dynamics from ergodic to non-ergodic is called emergent ergodicity breaking that, in the present case, emerges due to the effect of Fano interference. In the non-ergodic case, the system prepared in the dark state $(\left|1\right> - \left|2\right>)/\sqrt{2}$ stays trapped in this state forever and the PDF for current reads $P(t,j) = \delta(j)$. A similar situation occurs when the system is prepared outside the dark state with the only difference that then the PDF for current $P(t,j)$ is nontrivial, with a non-zero average current and a Gaussian-like shape similar to the dotted and dot-dashed PDFs shown in Figs.~\ref{fig:PDF} e) and f). 

The shape of the scaled cumulant generating function at the critical point and in its vicinity can be understood using relatively simple arguments (see Sec.\ 6.2 in Ref.~\cite{Manzano2018}; we also refer to Ref.~\cite{Vroylandt2019}, where the interested reader may found further information about large deviations in systems exhibiting emergent ergodicity breaking). In case of multiple steady-states with different stationary currents, the scaled cumulant generating function can be up to the first order in $u$ written as 
\begin{equation}
\lambda(u) \approx
\begin{cases}
- J_{\rm max} u, \quad u < 0,\\
- J_{\rm min} u, \quad u > 0,
\end{cases}
\end{equation}
where $J_{\rm max}$ is the largest possible value of the average steady-state current and $J_{\rm min}$ is the smallest one. In our case, the smallest current is $J_{\rm min} = 0$ and it corresponds to the excitation trapped in the dark state. The largest current $J_{\rm max}$ is then obtained for excitations running through the other steady state. Its value in the parameter regime $n_h \gg 1 \gg n_m$ is given by Eq.~\eqref{eq:quantumJ}. For the parameters taken for the full line in Fig.~\ref{fig:PDF} b), $J_{\rm max} \approx 6.55$. The behavior of the scaled cumulant generating function at $u\approx 3$, symmetric to that for $u \approx 0$, follows from the Gallavotti--Cohen fluctuation theorem~\cite{Manzano2018}
\begin{equation}
I(j) - I(-j) = \frac{\sigma}{k_{\rm B}}  = \frac{c j}{k_{\rm B}},
\label{eq:I}
\end{equation}
where 
\begin{equation}
c = \left(- \frac{E_4-E_3}{T_c} -\frac{E_1-E_4}{T_h} + \frac{E_1-E_3}{T_m}\right)
\end{equation}
can be read out from Eq.~\eqref{eq:sigma}. The fluctuation theorem \eqref{eq:I} is a consequence of microscopic time reversibility that implies that the operator $\mathcal{R}$ in Eq.~\eqref{eq:ME} obeys the local detailed balance condition~\cite{Manzano2018}. The theorem can be also obtained from the standard steady-state fluctuation theorem for entropy production $P(\sigma)/P(-\sigma) = \exp(-\sigma/k_{\rm B})$ \cite{Seifert2008}. 

Rewriting the fluctuation theorem \eqref{eq:I} in terms of scaled cumulant generating function yields~\cite{Manzano2018}
\begin{equation}
\lambda(u) = \lambda(c-u),
\end{equation}
which implies that near the point $c$ we can write 
\begin{equation}
\lambda(u) \approx
\begin{cases}
J_{\rm min} (u - c), \quad u < c,\\
J_{\rm max} (u-c), \quad u > c.
\end{cases}
\end{equation}
For the parameters taken in the figures, we obtain $c \approx 2.95$ in perfect agreement with our numerical results.

\subsection{Consequences for refrigeration}  

The broad current PDFs near the critical point and thus near the maximum current regime show that the maximum current is rather unstable compared to smaller currents obtained further away from the critical point.  This suggests that the studied absorption refrigerator exhibits a certain trade-off between the stability of the cooling flux and its magnitude. 

In the classical regime, this trade-off can be overcome if the system operates in the parameter regime $\gamma_{12h} = \gamma_{1h} = \gamma_{2h}$, when the dark- state $(\left|1\right> - \left|2\right>)/\sqrt{2}$ decouples from the rest of the system, see Fig.~\ref{fig:system_fridge}. In such a case, the system prepared at the initial time in a state within the effective three-level system $(\left|1\right> + \left|2\right>)/\sqrt{2}$ -- $\left|3\right>$ -- $\left|4\right>$ can yield the maximum current without paying  
by large fluctuations caused by trajectories captured for extensive periods of time in the dark state. However, this parameter regime is very narrow and, for all other parameters, the fluctuations in the classical case are significantly larger than in the quantum regime, compare panels b) in Figs.~\ref{fig:shape1} and \ref{fig:shape2}.

The best operation regime of the studied absorption refrigerator is thus the classical regime with the decoupled dark state and the system initially prepared outside the dark state. In such a situation, the effect of noise-induced coherence effectively transforms the four-level system $\{\left|1\right>,\left|2\right>,\left|3\right>,\left|4\right>\}$, depicted in Fig.~\ref{fig:fridges}, into the three-level system $\{(\left|1\right> + \left|2\right>)/\sqrt{2},\left|3\right>,\left|4\right>\}$, depicted in Fig.~\ref{fig:system_fridge}, with doubled transition rates into/from the uppermost level $(\left|1\right> + \left|2\right>)/\sqrt{2}$. This doubling of transition rates is the reason for the increase in the mean current compared to the classical four-level situation (without coherence coupled to populations in Eq.~\eqref{eq:freq}).

 Nevertheless, preparing the system exactly with the parameters leading to the decoupled dark state would be very difficult since they correspond to a single point in the two-dimensional parameter space $\gamma_{12x}$, $x=m,h$. If the system preparation is not so well under control, which is a standard situation in experiments, the next best option is to operate the system in the quantum regime that exhibits much smaller fluctuations than the classical one without the fine-tuned parameters.

\subsection{Further remarks}  

Let us close this section with two remarks. (i) We checked that the current fluctuations in the considered system fulfill the standard thermodynamic uncertainty relation $(\left<j^2\right>-J^2)/J^2 \ge 2k_{\rm B}/(\sigma t)$ introduced in Refs.~\cite{Gingrich2016,Proesmans2017} and hence also the weaker relation $(\left<j^2\right>-J^2)/J^2 \ge 2/[\exp(k_{\rm B}/\sigma t) - 1]$ recently derived from the fluctuation theorem \cite{Hasegawa2019}. See also Refs.~\cite{Agarwalla2018,Macieszczak2018,Brandner2018,Miller2018} for further studies on thermodynamic uncertainty relations for quantum systems. (ii) We have tested that large deviation functions for the integrated currents
\begin{equation}
j_h = \frac{1}{t}\int_0^t dt' [n_{4\to 1}(t') - n_{1\to 4}(t')
+ n_{4\to 2}(t') - n_{2\to 4}(t')]
\label{eq:fluctjh}
\end{equation}
and
\begin{equation}
j_m = \frac{1}{t}\int_0^t dt' [n_{3\to 1}(t') - n_{1\to 3}(t')
+ n_{3\to 2}(t') - n_{2\to 3}(t')],
\label{eq:fluctjm}
\end{equation}
corresponding to the transitions caused by hot and cold reservoir, respectively, are the same as the large deviation function for the current~\eqref{eq:fluctj}, as predicted in Ref.~\cite{Kindermann2004}. This is an intuitive result since the integrated currents can increase/decrease only if the excitation runs a full counterclockwise/clockwise circle $\left|4\right> \leftrightarrow \left|1\right>$ \& $\left|2\right> \leftrightarrow \left|3\right> \leftrightarrow \left|4\right>$. The tilted matrices $\tilde{\mathcal{R}}_h(u)$ and $\tilde{\mathcal{R}}_m(u)$, needed for computation of the large deviation functions for currents ~\eqref{eq:fluctjh} and \eqref{eq:fluctjm}, can again be constructed in an intuitive manner by tilting suitable elements of the matrix $\mathcal{R}$. 

The only difference from the simple \emph{classical} case of the current~\eqref{eq:fluctj} is that now the excitation can travel not only to/from the individual states $\left|1\right>$ (described by $\rho_{11}$) and $\left|2\right>$ (described by $\rho_{22}$) but also their coherent combination (described by $\rho_R$). In case of the current~\eqref{eq:fluctjh}, we thus have to multiply by $\exp(-u)$ (positive increment of the current) the elements $\tilde{\gamma}_{1h}n_h^+$, $\tilde{\gamma}_{2h}n_h^+$, and $\tilde{\gamma}_{12h}n_h^+$ in the fourth column of the matrix ${\mathcal R}$ corresponding to transitions from state $\left|4\right>$ to states $\left|1\right>$, $\left|2\right>$, and their coherent combination, respectively. The elements corresponding to decrease of the current, that have to be multiplied by $\exp(u)$, are than $\tilde{\gamma}_{1h}n_h^-$, $\tilde{\gamma}_{2h}n_h^-$ and $\tilde{\gamma}_{12h}n_h^-$ in the fourth line of the matrix ${\mathcal R}$. Similar considerations imply that in the matrix $\tilde{\mathcal{R}}_m(u)$ the elements $\tilde{\gamma}_{1m}n_m^-$, $\tilde{\gamma}_{2h}n_m^-$ and $\tilde{\gamma}_{12h}n_m^-$ in the third line, corresponding to decrease in the current~\eqref{eq:fluctjm}, must be multiplied by $\exp(u)$ and the elements $\tilde{\gamma}_{1h}n_h^+$, $\tilde{\gamma}_{2h}n_h^+$, and $\tilde{\gamma}_{12h}n_h^+$ in the third column, corresponding to increase of the current~\eqref{eq:fluctjm}, must be multiplied by $\exp(-u)$.

\section{Conclusion}\label{sec:Concl}  

Our investigation of current fluctuations in a simple model of absorption refrigerator with degenerate energy levels reveals that the effect of noise-induced coherence can be used as a textbook example of a system exhibiting dynamic phase transitions in current statistics. For symmetric transition rates in the quantum optical master equation describing the system dynamics, the steady state attained by the system at a long time becomes non-unique and this leads to non-analytical scaled cumulant generating function for current, broad probability distributions for current, and thus also large current fluctuations. Interestingly, the refrigerator delivers the largest cooling flux for parameters in the vicinity of this symmetric parameter regime and thus the presence of the phase transition leading to large fluctuations negatively influences its performance. This effect is reminiscent of large power fluctuations~\cite{Holubec2017a} in heat engines operating in the vicinity of a phase transition~\cite{Campisi2016}. However, while the phase transition in Ref.~\cite{Campisi2016} is attained in the limit of many interacting heat engines and the corresponding fluctuations can be tamed by using a suitable fine-tuned driving protocol~\cite{Holubec2018b}, the phase transition in the present model occurs on the level of a single system and, for given parameters of the model, it can not be avoided.

The phase transition occurs only in the parameter regime that allows for a classical description of the system using a suitably rotated basis of the degenerate subspace. For parameters at the vicinity of the critical point, the current fluctuations are qualitatively the same for this classical regime and also for the remaining parameters composing what we call quantum regime. The main difference is the magnitude of fluctuations which is significantly larger in the classical case. The classical parameter regime offers only a narrow path to avoid these large fluctuations by a suitable initial preparation of the system in the decoupled subspace that yields the largest current. However, since the two subspaces completely decouple for a single point in the parameter space only, such a fine-tuning of the system dynamics seems unaccessible from the experimental point of view.

Because the degenerate steady state formed in the classical case is composed of two parts, the system behavior in the vicinity of the critical point can be described by a dichotomous process. This allowed us to provide an analytical description of the average current as well as a qualitative understanding of its fluctuations both in the classical and the quantum case, revealing the main differences between these two parameter regimes. 

Although we performed our investigation of the effect of noise-induced coherence using a specific model of absorption refrigerator, the main conclusions about the system behavior in the classical and quantum regime of operation presented here, and also in our previous study~\cite{Holubec2018}, are valid for arbitrary systems with degenerate energy levels in the spectrum that are weakly coupled to several (bosonic) thermal bath by dipole interactions and thus they fulfill assumptions used for the derivation of the quantum optical master equation.

\begin{acknowledgments}
This work was supported by the Czech Science Foundation (project No. 17-06716S). VH in addition gratefully acknowledges the support by the Alexander von Humboldt foundation.
\end{acknowledgments}



\begin{thebibliography}{49}%
\makeatletter
\providecommand \@ifxundefined [1]{%
 \@ifx{#1\undefined}
}%
\providecommand \@ifnum [1]{%
 \ifnum #1\expandafter \@firstoftwo
 \else \expandafter \@secondoftwo
 \fi
}%
\providecommand \@ifx [1]{%
 \ifx #1\expandafter \@firstoftwo
 \else \expandafter \@secondoftwo
 \fi
}%
\providecommand \natexlab [1]{#1}%
\providecommand \enquote  [1]{``#1''}%
\providecommand \bibnamefont  [1]{#1}%
\providecommand \bibfnamefont [1]{#1}%
\providecommand \citenamefont [1]{#1}%
\providecommand \href@noop [0]{\@secondoftwo}%
\providecommand \href [0]{\begingroup \@sanitize@url \@href}%
\providecommand \@href[1]{\@@startlink{#1}\@@href}%
\providecommand \@@href[1]{\endgroup#1\@@endlink}%
\providecommand \@sanitize@url [0]{\catcode `\\12\catcode `\$12\catcode
  `\&12\catcode `\#12\catcode `\^12\catcode `\_12\catcode `\%12\relax}%
\providecommand \@@startlink[1]{}%
\providecommand \@@endlink[0]{}%
\providecommand \url  [0]{\begingroup\@sanitize@url \@url }%
\providecommand \@url [1]{\endgroup\@href {#1}{\urlprefix }}%
\providecommand \urlprefix  [0]{URL }%
\providecommand \Eprint [0]{\href }%
\providecommand \doibase [0]{http://dx.doi.org/}%
\providecommand \selectlanguage [0]{\@gobble}%
\providecommand \bibinfo  [0]{\@secondoftwo}%
\providecommand \bibfield  [0]{\@secondoftwo}%
\providecommand \translation [1]{[#1]}%
\providecommand \BibitemOpen [0]{}%
\providecommand \bibitemStop [0]{}%
\providecommand \bibitemNoStop [0]{.\EOS\space}%
\providecommand \EOS [0]{\spacefactor3000\relax}%
\providecommand \BibitemShut  [1]{\csname bibitem#1\endcsname}%
\let\auto@bib@innerbib\@empty
\bibitem [{\citenamefont {Dorfman}\ \emph {et~al.}(2013)\citenamefont
  {Dorfman}, \citenamefont {Voronine}, \citenamefont {Mukamel},\ and\
  \citenamefont {Scully}}]{Dorfman2013}%
  \BibitemOpen
  \bibfield  {author} {\bibinfo {author} {\bibfnamefont {K.~E.}\ \bibnamefont
  {Dorfman}}, \bibinfo {author} {\bibfnamefont {D.~V.}\ \bibnamefont
  {Voronine}}, \bibinfo {author} {\bibfnamefont {S.}~\bibnamefont {Mukamel}}, \
  and\ \bibinfo {author} {\bibfnamefont {M.~O.}\ \bibnamefont {Scully}},\
  }\href {\doibase 10.1073/pnas.1212666110} {\bibfield  {journal} {\bibinfo
  {journal} {Proc. Natl. Acad. Sci. U. S. A.}\ }\textbf {\bibinfo {volume}
  {110}},\ \bibinfo {pages} {2746} (\bibinfo {year} {2013})}\BibitemShut
  {NoStop}%
\bibitem [{\citenamefont {Del~Campo}, \citenamefont {Goold},\ and\
  \citenamefont {Paternostro}(2014)}]{DelCampo2014}%
  \BibitemOpen
  \bibfield  {author} {\bibinfo {author} {\bibfnamefont {A.}~\bibnamefont
  {Del~Campo}}, \bibinfo {author} {\bibfnamefont {J.}~\bibnamefont {Goold}}, \
  and\ \bibinfo {author} {\bibfnamefont {M.}~\bibnamefont {Paternostro}},\
  }\href@noop {} {\bibfield  {journal} {\bibinfo  {journal} {Sci. Rep.}\
  }\textbf {\bibinfo {volume} {4}},\ \bibinfo {pages} {6208} (\bibinfo {year}
  {2014})}\BibitemShut {NoStop}%
\bibitem [{\citenamefont {Vinjanampathy}\ and\ \citenamefont
  {Anders}(2016)}]{Vinjanampathy2016}%
  \BibitemOpen
  \bibfield  {author} {\bibinfo {author} {\bibfnamefont {S.}~\bibnamefont
  {Vinjanampathy}}\ and\ \bibinfo {author} {\bibfnamefont {J.}~\bibnamefont
  {Anders}},\ }\href {\doibase 10.1080/00107514.2016.1201896} {\bibfield
  {journal} {\bibinfo  {journal} {Contemp. Phys.}\ }\textbf {\bibinfo {volume}
  {57}},\ \bibinfo {pages} {545} (\bibinfo {year} {2016})}\BibitemShut
  {NoStop}%
\bibitem [{\citenamefont {Ghosh}\ \emph {et~al.}(2019)\citenamefont {Ghosh},
  \citenamefont {Mukherjee}, \citenamefont {Niedenzu},\ and\ \citenamefont
  {Kurizki}}]{Ghosh2019}%
  \BibitemOpen
  \bibfield  {author} {\bibinfo {author} {\bibfnamefont {A.}~\bibnamefont
  {Ghosh}}, \bibinfo {author} {\bibfnamefont {V.}~\bibnamefont {Mukherjee}},
  \bibinfo {author} {\bibfnamefont {W.}~\bibnamefont {Niedenzu}}, \ and\
  \bibinfo {author} {\bibfnamefont {G.}~\bibnamefont {Kurizki}},\ }\href
  {\doibase 10.1140/epjst/e2019-800060-7} {\bibfield  {journal} {\bibinfo
  {journal} {Eur. Phys. J. Spec. Tops.}\ }\textbf {\bibinfo {volume} {227}},\
  \bibinfo {pages} {2043} (\bibinfo {year} {2019})}\BibitemShut {NoStop}%
\bibitem [{\citenamefont {Binder}\ \emph {et~al.}(2019)\citenamefont {Binder},
  \citenamefont {Correa}, \citenamefont {Gogolin}, \citenamefont {Anders},\
  and\ \citenamefont {Adesso}}]{Binder2019}%
  \BibitemOpen
  \bibinfo {editor} {\bibfnamefont {F.}~\bibnamefont {Binder}}, \bibinfo
  {editor} {\bibfnamefont {L.~A.}\ \bibnamefont {Correa}}, \bibinfo {editor}
  {\bibfnamefont {C.}~\bibnamefont {Gogolin}}, \bibinfo {editor} {\bibfnamefont
  {J.}~\bibnamefont {Anders}}, \ and\ \bibinfo {editor} {\bibfnamefont
  {G.}~\bibnamefont {Adesso}},\ eds.,\ \href@noop {} {\emph {\bibinfo {title}
  {Thermodynamics in the Quantum Regime: Fundamental Aspects and New
  Directions}}}\ (\bibinfo  {publisher} {Springer International},\ \bibinfo
  {year} {2019})\BibitemShut {NoStop}%
\bibitem [{\citenamefont {Holubec}\ and\ \citenamefont
  {Novotn{\'y}}(2018)}]{Holubec2018}%
  \BibitemOpen
  \bibfield  {author} {\bibinfo {author} {\bibfnamefont {V.}~\bibnamefont
  {Holubec}}\ and\ \bibinfo {author} {\bibfnamefont {T.}~\bibnamefont
  {Novotn{\'y}}},\ }\href {\doibase 10.1007/s10909-018-1960-x} {\bibfield
  {journal} {\bibinfo  {journal} {J. Low Temp. Phys.}\ }\textbf {\bibinfo
  {volume} {192}},\ \bibinfo {pages} {147} (\bibinfo {year}
  {2018})}\BibitemShut {NoStop}%
\bibitem [{\citenamefont {Gonz{\'a}lez}\ \emph {et~al.}(2019)\citenamefont
  {Gonz{\'a}lez}, \citenamefont {Palao}, \citenamefont {Alonso},\ and\
  \citenamefont {Correa}}]{Gonzalez2019}%
  \BibitemOpen
  \bibfield  {author} {\bibinfo {author} {\bibfnamefont {J.~O.}\ \bibnamefont
  {Gonz{\'a}lez}}, \bibinfo {author} {\bibfnamefont {J.}~\bibnamefont {Palao}},
  \bibinfo {author} {\bibfnamefont {D.}~\bibnamefont {Alonso}}, \ and\ \bibinfo
  {author} {\bibfnamefont {L.~A.}\ \bibnamefont {Correa}},\ }\href {\doibase
  10.1103/PhysRevE.99.062102} {\bibfield  {journal} {\bibinfo  {journal} {Phys.
  Rev. E}\ }\textbf {\bibinfo {volume} {99}},\ \bibinfo {pages} {062102}
  (\bibinfo {year} {2019})}\BibitemShut {NoStop}%
\bibitem [{\citenamefont {Scully}\ \emph {et~al.}(2011)\citenamefont {Scully},
  \citenamefont {Chapin}, \citenamefont {Dorfman}, \citenamefont {Kim},\ and\
  \citenamefont {Svidzinsky}}]{Scully2011}%
  \BibitemOpen
  \bibfield  {author} {\bibinfo {author} {\bibfnamefont {M.~O.}\ \bibnamefont
  {Scully}}, \bibinfo {author} {\bibfnamefont {K.~R.}\ \bibnamefont {Chapin}},
  \bibinfo {author} {\bibfnamefont {K.~E.}\ \bibnamefont {Dorfman}}, \bibinfo
  {author} {\bibfnamefont {M.~B.}\ \bibnamefont {Kim}}, \ and\ \bibinfo
  {author} {\bibfnamefont {A.}~\bibnamefont {Svidzinsky}},\ }\href {\doibase
  10.1073/pnas.1110234108} {\bibfield  {journal} {\bibinfo  {journal} {Proc.
  Natl. Acad. Sci. U. S. A.}\ }\textbf {\bibinfo {volume} {108}},\ \bibinfo
  {pages} {15097} (\bibinfo {year} {2011})}\BibitemShut {NoStop}%
\bibitem [{\citenamefont {Svidzinsky}, \citenamefont {Dorfman},\ and\
  \citenamefont {Scully}(2011)}]{Svidzinsky2011}%
  \BibitemOpen
  \bibfield  {author} {\bibinfo {author} {\bibfnamefont {A.~A.}\ \bibnamefont
  {Svidzinsky}}, \bibinfo {author} {\bibfnamefont {K.~E.}\ \bibnamefont
  {Dorfman}}, \ and\ \bibinfo {author} {\bibfnamefont {M.~O.}\ \bibnamefont
  {Scully}},\ }\href {\doibase 10.1103/PhysRevA.84.053818} {\bibfield
  {journal} {\bibinfo  {journal} {Phys. Rev. A}\ }\textbf {\bibinfo {volume}
  {84}},\ \bibinfo {pages} {053818} (\bibinfo {year} {2011})}\BibitemShut
  {NoStop}%
\bibitem [{\citenamefont {Svidzinsky}, \citenamefont {Dorfman},\ and\
  \citenamefont {Scully}(2012)}]{Svidzinsky2012}%
  \BibitemOpen
  \bibfield  {author} {\bibinfo {author} {\bibfnamefont {A.}~\bibnamefont
  {Svidzinsky}}, \bibinfo {author} {\bibfnamefont {K.}~\bibnamefont {Dorfman}},
  \ and\ \bibinfo {author} {\bibfnamefont {M.}~\bibnamefont {Scully}},\ }\href
  {\doibase doi:10.2478/coph-2012-0002} {\bibfield  {journal} {\bibinfo
  {journal} {Coherent Optical Phenomena}\ }\textbf {\bibinfo {volume} {1}},\
  \bibinfo {pages} {7} (\bibinfo {year} {2012})}\BibitemShut {NoStop}%
\bibitem [{\citenamefont {Fano}(1961)}]{Fano1961}%
  \BibitemOpen
  \bibfield  {author} {\bibinfo {author} {\bibfnamefont {U.}~\bibnamefont
  {Fano}},\ }\href {\doibase 10.1103/PhysRev.124.1866} {\bibfield  {journal}
  {\bibinfo  {journal} {Phys. Rev.}\ }\textbf {\bibinfo {volume} {124}},\
  \bibinfo {pages} {1866} (\bibinfo {year} {1961})}\BibitemShut {NoStop}%
\bibitem [{\citenamefont {Muralidharan}\ and\ \citenamefont
  {Datta}(2007)}]{Muralidharan2007}%
  \BibitemOpen
  \bibfield  {author} {\bibinfo {author} {\bibfnamefont {B.}~\bibnamefont
  {Muralidharan}}\ and\ \bibinfo {author} {\bibfnamefont {S.}~\bibnamefont
  {Datta}},\ }\href {\doibase 10.1103/PhysRevB.76.035432} {\bibfield  {journal}
  {\bibinfo  {journal} {Phys. Rev. B}\ }\textbf {\bibinfo {volume} {76}},\
  \bibinfo {pages} {035432} (\bibinfo {year} {2007})}\BibitemShut {NoStop}%
\bibitem [{\citenamefont {Harris}(1989)}]{Harris1989}%
  \BibitemOpen
  \bibfield  {author} {\bibinfo {author} {\bibfnamefont {S.~E.}\ \bibnamefont
  {Harris}},\ }\href {\doibase 10.1103/PhysRevLett.62.1033} {\bibfield
  {journal} {\bibinfo  {journal} {Phys. Rev. Lett.}\ }\textbf {\bibinfo
  {volume} {62}},\ \bibinfo {pages} {1033} (\bibinfo {year}
  {1989})}\BibitemShut {NoStop}%
\bibitem [{\citenamefont {Boller}, \citenamefont {Imamo\ifmmode~\breve{g}\else
  \u{g}\fi{}lu},\ and\ \citenamefont {Harris}(1991)}]{Boller1991}%
  \BibitemOpen
  \bibfield  {author} {\bibinfo {author} {\bibfnamefont {K.-J.}\ \bibnamefont
  {Boller}}, \bibinfo {author} {\bibfnamefont {A.}~\bibnamefont
  {Imamo\ifmmode~\breve{g}\else \u{g}\fi{}lu}}, \ and\ \bibinfo {author}
  {\bibfnamefont {S.~E.}\ \bibnamefont {Harris}},\ }\href {\doibase
  10.1103/PhysRevLett.66.2593} {\bibfield  {journal} {\bibinfo  {journal}
  {Phys. Rev. Lett.}\ }\textbf {\bibinfo {volume} {66}},\ \bibinfo {pages}
  {2593} (\bibinfo {year} {1991})}\BibitemShut {NoStop}%
\bibitem [{\citenamefont {Rybin}\ \emph {et~al.}(2015)\citenamefont {Rybin},
  \citenamefont {Filonov}, \citenamefont {Belov}, \citenamefont {Kivshar},\
  and\ \citenamefont {Limonov}}]{Rybin2015}%
  \BibitemOpen
  \bibfield  {author} {\bibinfo {author} {\bibfnamefont {M.~V.}\ \bibnamefont
  {Rybin}}, \bibinfo {author} {\bibfnamefont {D.~S.}\ \bibnamefont {Filonov}},
  \bibinfo {author} {\bibfnamefont {P.~A.}\ \bibnamefont {Belov}}, \bibinfo
  {author} {\bibfnamefont {Y.~S.}\ \bibnamefont {Kivshar}}, \ and\ \bibinfo
  {author} {\bibfnamefont {M.~F.}\ \bibnamefont {Limonov}},\ }\href@noop {}
  {\bibfield  {journal} {\bibinfo  {journal} {Sci. Rep.}\ }\textbf {\bibinfo
  {volume} {5}},\ \bibinfo {pages} {8774} (\bibinfo {year} {2015})}\BibitemShut
  {NoStop}%
\bibitem [{\citenamefont {Joe}, \citenamefont {Satanin},\ and\ \citenamefont
  {Kim}(2006)}]{Joe2006}%
  \BibitemOpen
  \bibfield  {author} {\bibinfo {author} {\bibfnamefont {Y.~S.}\ \bibnamefont
  {Joe}}, \bibinfo {author} {\bibfnamefont {A.~M.}\ \bibnamefont {Satanin}}, \
  and\ \bibinfo {author} {\bibfnamefont {C.~S.}\ \bibnamefont {Kim}},\ }\href
  {\doibase 10.1088/0031-8949/74/2/020} {\bibfield  {journal} {\bibinfo
  {journal} {Phys. Scr.}\ }\textbf {\bibinfo {volume} {74}},\ \bibinfo {pages}
  {259} (\bibinfo {year} {2006})}\BibitemShut {NoStop}%
\bibitem [{\citenamefont {Creatore}\ \emph {et~al.}(2013)\citenamefont
  {Creatore}, \citenamefont {Parker}, \citenamefont {Emmott},\ and\
  \citenamefont {Chin}}]{Creatore2013}%
  \BibitemOpen
  \bibfield  {author} {\bibinfo {author} {\bibfnamefont {C.}~\bibnamefont
  {Creatore}}, \bibinfo {author} {\bibfnamefont {M.~A.}\ \bibnamefont
  {Parker}}, \bibinfo {author} {\bibfnamefont {S.}~\bibnamefont {Emmott}}, \
  and\ \bibinfo {author} {\bibfnamefont {A.~W.}\ \bibnamefont {Chin}},\ }\href
  {\doibase 10.1103/PhysRevLett.111.253601} {\bibfield  {journal} {\bibinfo
  {journal} {Phys. Rev. Lett.}\ }\textbf {\bibinfo {volume} {111}},\ \bibinfo
  {pages} {253601} (\bibinfo {year} {2013})}\BibitemShut {NoStop}%
\bibitem [{\citenamefont {Chen}, \citenamefont {Chiu},\ and\ \citenamefont
  {Chen}(2016)}]{Chen2016}%
  \BibitemOpen
  \bibfield  {author} {\bibinfo {author} {\bibfnamefont {H.-B.}\ \bibnamefont
  {Chen}}, \bibinfo {author} {\bibfnamefont {P.-Y.}\ \bibnamefont {Chiu}}, \
  and\ \bibinfo {author} {\bibfnamefont {Y.-N.}\ \bibnamefont {Chen}},\ }\href
  {\doibase 10.1103/PhysRevE.94.052101} {\bibfield  {journal} {\bibinfo
  {journal} {Phys. Rev. E}\ }\textbf {\bibinfo {volume} {94}},\ \bibinfo
  {pages} {052101} (\bibinfo {year} {2016})}\BibitemShut {NoStop}%
\bibitem [{\citenamefont {Manzano}\ and\ \citenamefont
  {Hurtado}(2018)}]{Manzano2018}%
  \BibitemOpen
  \bibfield  {author} {\bibinfo {author} {\bibfnamefont {D.}~\bibnamefont
  {Manzano}}\ and\ \bibinfo {author} {\bibfnamefont {P.}~\bibnamefont
  {Hurtado}},\ }\href {\doibase 10.1080/00018732.2018.1519981} {\bibfield
  {journal} {\bibinfo  {journal} {Adv. Phys.}\ }\textbf {\bibinfo {volume}
  {67}},\ \bibinfo {pages} {1} (\bibinfo {year} {2018})}\BibitemShut {NoStop}%
\bibitem [{\citenamefont {Correa}\ \emph {et~al.}(2013)\citenamefont {Correa},
  \citenamefont {Palao}, \citenamefont {Adesso},\ and\ \citenamefont
  {Alonso}}]{Correa2013}%
  \BibitemOpen
  \bibfield  {author} {\bibinfo {author} {\bibfnamefont {L.~A.}\ \bibnamefont
  {Correa}}, \bibinfo {author} {\bibfnamefont {J.~P.}\ \bibnamefont {Palao}},
  \bibinfo {author} {\bibfnamefont {G.}~\bibnamefont {Adesso}}, \ and\ \bibinfo
  {author} {\bibfnamefont {D.}~\bibnamefont {Alonso}},\ }\href {\doibase
  10.1103/PhysRevE.87.042131} {\bibfield  {journal} {\bibinfo  {journal} {Phys.
  Rev. E}\ }\textbf {\bibinfo {volume} {87}},\ \bibinfo {pages} {042131}
  (\bibinfo {year} {2013})}\BibitemShut {NoStop}%
\bibitem [{\citenamefont {Brask}\ and\ \citenamefont
  {Brunner}(2015)}]{Brask2015}%
  \BibitemOpen
  \bibfield  {author} {\bibinfo {author} {\bibfnamefont {J.~B.}\ \bibnamefont
  {Brask}}\ and\ \bibinfo {author} {\bibfnamefont {N.}~\bibnamefont
  {Brunner}},\ }\href {\doibase 10.1103/PhysRevE.92.062101} {\bibfield
  {journal} {\bibinfo  {journal} {Phys. Rev. E}\ }\textbf {\bibinfo {volume}
  {92}},\ \bibinfo {pages} {062101} (\bibinfo {year} {2015})}\BibitemShut
  {NoStop}%
\bibitem [{\citenamefont {Correa}\ \emph {et~al.}(2014)\citenamefont {Correa},
  \citenamefont {Palao}, \citenamefont {Alonso},\ and\ \citenamefont
  {Adesso}}]{Correa2014}%
  \BibitemOpen
  \bibfield  {author} {\bibinfo {author} {\bibfnamefont {L.~A.}\ \bibnamefont
  {Correa}}, \bibinfo {author} {\bibfnamefont {J.}~\bibnamefont {Palao}},
  \bibinfo {author} {\bibfnamefont {D.}~\bibnamefont {Alonso}}, \ and\ \bibinfo
  {author} {\bibfnamefont {G.}~\bibnamefont {Adesso}},\ }\href {\doibase
  10.1038/srep03949} {\bibfield  {journal} {\bibinfo  {journal} {Sci. Rep.}\
  }\textbf {\bibinfo {volume} {4}},\ \bibinfo {pages} {3949} (\bibinfo {year}
  {2014})}\BibitemShut {NoStop}%
\bibitem [{\citenamefont {Silva}, \citenamefont {Skrzypczyk},\ and\
  \citenamefont {Brunner}(2015)}]{Silva2015}%
  \BibitemOpen
  \bibfield  {author} {\bibinfo {author} {\bibfnamefont {R.}~\bibnamefont
  {Silva}}, \bibinfo {author} {\bibfnamefont {P.}~\bibnamefont {Skrzypczyk}}, \
  and\ \bibinfo {author} {\bibfnamefont {N.}~\bibnamefont {Brunner}},\ }\href
  {\doibase 10.1103/PhysRevE.92.012136} {\bibfield  {journal} {\bibinfo
  {journal} {Phys. Rev. E}\ }\textbf {\bibinfo {volume} {92}},\ \bibinfo
  {pages} {012136} (\bibinfo {year} {2015})}\BibitemShut {NoStop}%
\bibitem [{\citenamefont {Gonz{\'a}lez}, \citenamefont {Palao},\ and\
  \citenamefont {Alonso}(2017)}]{Gonzalez2017}%
  \BibitemOpen
  \bibfield  {author} {\bibinfo {author} {\bibfnamefont {J.~O.}\ \bibnamefont
  {Gonz{\'a}lez}}, \bibinfo {author} {\bibfnamefont {J.~P.}\ \bibnamefont
  {Palao}}, \ and\ \bibinfo {author} {\bibfnamefont {D.}~\bibnamefont
  {Alonso}},\ }\href@noop {} {\bibfield  {journal} {\bibinfo  {journal} {New J.
  Phys.}\ }\textbf {\bibinfo {volume} {19}},\ \bibinfo {pages} {113037}
  (\bibinfo {year} {2017})}\BibitemShut {NoStop}%
\bibitem [{\citenamefont {Segal}(2018)}]{Segal2018}%
  \BibitemOpen
  \bibfield  {author} {\bibinfo {author} {\bibfnamefont {D.}~\bibnamefont
  {Segal}},\ }\href {\doibase 10.1103/PhysRevE.97.052145} {\bibfield  {journal}
  {\bibinfo  {journal} {Phys. Rev. E}\ }\textbf {\bibinfo {volume} {97}},\
  \bibinfo {pages} {052145} (\bibinfo {year} {2018})}\BibitemShut {NoStop}%
\bibitem [{\citenamefont {Kilgour}\ and\ \citenamefont
  {Segal}(2018)}]{Kilgour2018}%
  \BibitemOpen
  \bibfield  {author} {\bibinfo {author} {\bibfnamefont {M.}~\bibnamefont
  {Kilgour}}\ and\ \bibinfo {author} {\bibfnamefont {D.}~\bibnamefont
  {Segal}},\ }\href {\doibase 10.1103/PhysRevE.98.012117} {\bibfield  {journal}
  {\bibinfo  {journal} {Phys. Rev. E}\ }\textbf {\bibinfo {volume} {98}},\
  \bibinfo {pages} {012117} (\bibinfo {year} {2018})}\BibitemShut {NoStop}%
\bibitem [{\citenamefont {Mitchison}(2019)}]{Mitchison2019}%
  \BibitemOpen
  \bibfield  {author} {\bibinfo {author} {\bibfnamefont {M.~T.}\ \bibnamefont
  {Mitchison}},\ }\href {\doibase 10.1080/00107514.2019.1631555} {\bibfield
  {journal} {\bibinfo  {journal} {Contemporary Physics}\ }\textbf {\bibinfo
  {volume} {60}},\ \bibinfo {pages} {164} (\bibinfo {year} {2019})}\BibitemShut
  {NoStop}%
\bibitem [{\citenamefont {Levy}\ and\ \citenamefont
  {Kosloff}(2012)}]{Levy2012}%
  \BibitemOpen
  \bibfield  {author} {\bibinfo {author} {\bibfnamefont {A.}~\bibnamefont
  {Levy}}\ and\ \bibinfo {author} {\bibfnamefont {R.}~\bibnamefont {Kosloff}},\
  }\href {\doibase 10.1103/PhysRevLett.108.070604} {\bibfield  {journal}
  {\bibinfo  {journal} {Phys. Rev. Lett.}\ }\textbf {\bibinfo {volume} {108}},\
  \bibinfo {pages} {070604} (\bibinfo {year} {2012})}\BibitemShut {NoStop}%
\bibitem [{\citenamefont {Holubec}\ and\ \citenamefont
  {Ryabov}(2018)}]{Holubec2018b}%
  \BibitemOpen
  \bibfield  {author} {\bibinfo {author} {\bibfnamefont {V.}~\bibnamefont
  {Holubec}}\ and\ \bibinfo {author} {\bibfnamefont {A.}~\bibnamefont
  {Ryabov}},\ }\href {\doibase 10.1103/PhysRevLett.121.120601} {\bibfield
  {journal} {\bibinfo  {journal} {Phys. Rev. Lett.}\ }\textbf {\bibinfo
  {volume} {121}},\ \bibinfo {pages} {120601} (\bibinfo {year}
  {2018})}\BibitemShut {NoStop}%
\bibitem [{\citenamefont {Pietzonka}\ and\ \citenamefont
  {Seifert}(2018)}]{Pietzonka2018}%
  \BibitemOpen
  \bibfield  {author} {\bibinfo {author} {\bibfnamefont {P.}~\bibnamefont
  {Pietzonka}}\ and\ \bibinfo {author} {\bibfnamefont {U.}~\bibnamefont
  {Seifert}},\ }\href {\doibase 10.1103/PhysRevLett.120.190602} {\bibfield
  {journal} {\bibinfo  {journal} {Phys. Rev. Lett.}\ }\textbf {\bibinfo
  {volume} {120}},\ \bibinfo {pages} {190602} (\bibinfo {year}
  {2018})}\BibitemShut {NoStop}%
\bibitem [{\citenamefont {Breuer}\ and\ \citenamefont
  {Petruccione}(2002)}]{Breuer2002}%
  \BibitemOpen
  \bibfield  {author} {\bibinfo {author} {\bibfnamefont {H.-P.}\ \bibnamefont
  {Breuer}}\ and\ \bibinfo {author} {\bibfnamefont {F.}~\bibnamefont
  {Petruccione}},\ }\href@noop {} {\emph {\bibinfo {title} {The theory of open
  quantum systems}}}\ (\bibinfo  {publisher} {Oxford University Press on
  Demand},\ \bibinfo {year} {2002})\BibitemShut {NoStop}%
\bibitem [{\citenamefont {Sedl{\'a}k}(2018)}]{Sedlak2018}%
  \BibitemOpen
  \bibfield  {author} {\bibinfo {author} {\bibfnamefont {O.}~\bibnamefont
  {Sedl{\'a}k}},\ }\href@noop {} {\enquote {\bibinfo {title} {Quantum
  thermodynamics},}\ } (\bibinfo {year} {2018}),\ \bibinfo {note} {{Bachelor
  thesis, Univerzita Karlova, Matematicko-fyzik{\'a}ln{\'\i}
  fakulta}}\BibitemShut {NoStop}%
\bibitem [{\citenamefont {Touchette}(2009)}]{Touchette2009}%
  \BibitemOpen
  \bibfield  {author} {\bibinfo {author} {\bibfnamefont {H.}~\bibnamefont
  {Touchette}},\ }\href {\doibase
  https://doi.org/10.1016/j.physrep.2009.05.002} {\bibfield  {journal}
  {\bibinfo  {journal} {Phys. Rep.}\ }\textbf {\bibinfo {volume} {478}},\
  \bibinfo {pages} {1 } (\bibinfo {year} {2009})}\BibitemShut {NoStop}%
\bibitem [{\citenamefont {Bertini}\ \emph {et~al.}(2001)\citenamefont
  {Bertini}, \citenamefont {De~Sole}, \citenamefont {Gabrielli}, \citenamefont
  {Jona-Lasinio},\ and\ \citenamefont {Landim}}]{Bertini2001}%
  \BibitemOpen
  \bibfield  {author} {\bibinfo {author} {\bibfnamefont {L.}~\bibnamefont
  {Bertini}}, \bibinfo {author} {\bibfnamefont {A.}~\bibnamefont {De~Sole}},
  \bibinfo {author} {\bibfnamefont {D.}~\bibnamefont {Gabrielli}}, \bibinfo
  {author} {\bibfnamefont {G.}~\bibnamefont {Jona-Lasinio}}, \ and\ \bibinfo
  {author} {\bibfnamefont {C.}~\bibnamefont {Landim}},\ }\href {\doibase
  10.1103/PhysRevLett.87.040601} {\bibfield  {journal} {\bibinfo  {journal}
  {Phys. Rev. Lett.}\ }\textbf {\bibinfo {volume} {87}},\ \bibinfo {pages}
  {040601} (\bibinfo {year} {2001})}\BibitemShut {NoStop}%
\bibitem [{\citenamefont {Holubec}, \citenamefont {Kroy},\ and\ \citenamefont
  {Steffenoni}(2019)}]{Holubec2018a}%
  \BibitemOpen
  \bibfield  {author} {\bibinfo {author} {\bibfnamefont {V.}~\bibnamefont
  {Holubec}}, \bibinfo {author} {\bibfnamefont {K.}~\bibnamefont {Kroy}}, \
  and\ \bibinfo {author} {\bibfnamefont {S.}~\bibnamefont {Steffenoni}},\
  }\href {\doibase 10.1103/PhysRevE.99.032117} {\bibfield  {journal} {\bibinfo
  {journal} {Phys. Rev. E}\ }\textbf {\bibinfo {volume} {99}},\ \bibinfo
  {pages} {032117} (\bibinfo {year} {2019})}\BibitemShut {NoStop}%
\bibitem [{\citenamefont {Bulnes~Cuetara}, \citenamefont {Esposito},\ and\
  \citenamefont {Schaller}(2016)}]{BulnesCuetara2016}%
  \BibitemOpen
  \bibfield  {author} {\bibinfo {author} {\bibfnamefont {G.}~\bibnamefont
  {Bulnes~Cuetara}}, \bibinfo {author} {\bibfnamefont {M.}~\bibnamefont
  {Esposito}}, \ and\ \bibinfo {author} {\bibfnamefont {G.}~\bibnamefont
  {Schaller}},\ }\href {\doibase 10.3390/e18120447} {\bibfield  {journal}
  {\bibinfo  {journal} {Entropy}\ }\textbf {\bibinfo {volume} {18}},\ \bibinfo
  {pages} {447} (\bibinfo {year} {2016})}\BibitemShut {NoStop}%
\bibitem [{\citenamefont {Garrahan}\ and\ \citenamefont
  {Lesanovsky}(2010)}]{Garrahan2010}%
  \BibitemOpen
  \bibfield  {author} {\bibinfo {author} {\bibfnamefont {J.~P.}\ \bibnamefont
  {Garrahan}}\ and\ \bibinfo {author} {\bibfnamefont {I.}~\bibnamefont
  {Lesanovsky}},\ }\href {\doibase 10.1103/PhysRevLett.104.160601} {\bibfield
  {journal} {\bibinfo  {journal} {Phys. Rev. Lett.}\ }\textbf {\bibinfo
  {volume} {104}},\ \bibinfo {pages} {160601} (\bibinfo {year}
  {2010})}\BibitemShut {NoStop}%
\bibitem [{\citenamefont {Vroylandt}\ and\ \citenamefont
  {Verley}(2019)}]{Vroylandt2019}%
  \BibitemOpen
  \bibfield  {author} {\bibinfo {author} {\bibfnamefont {H.}~\bibnamefont
  {Vroylandt}}\ and\ \bibinfo {author} {\bibfnamefont {G.}~\bibnamefont
  {Verley}},\ }\href {\doibase 10.1007/s10955-018-2186-7} {\bibfield  {journal}
  {\bibinfo  {journal} {J. Stat. Phys.}\ }\textbf {\bibinfo {volume} {174}},\
  \bibinfo {pages} {404} (\bibinfo {year} {2019})}\BibitemShut {NoStop}%
\bibitem [{\citenamefont {Seifert}(2008)}]{Seifert2008}%
  \BibitemOpen
  \bibfield  {author} {\bibinfo {author} {\bibfnamefont {U.}~\bibnamefont
  {Seifert}},\ }\href {\doibase 10.1140/epjb/e2008-00001-9} {\bibfield
  {journal} {\bibinfo  {journal} {Eur. Phys. J. B}\ }\textbf {\bibinfo {volume}
  {64}},\ \bibinfo {pages} {423} (\bibinfo {year} {2008})}\BibitemShut
  {NoStop}%
\bibitem [{\citenamefont {Gingrich}\ \emph {et~al.}(2016)\citenamefont
  {Gingrich}, \citenamefont {Horowitz}, \citenamefont {Perunov},\ and\
  \citenamefont {England}}]{Gingrich2016}%
  \BibitemOpen
  \bibfield  {author} {\bibinfo {author} {\bibfnamefont {T.~R.}\ \bibnamefont
  {Gingrich}}, \bibinfo {author} {\bibfnamefont {J.~M.}\ \bibnamefont
  {Horowitz}}, \bibinfo {author} {\bibfnamefont {N.}~\bibnamefont {Perunov}}, \
  and\ \bibinfo {author} {\bibfnamefont {J.~L.}\ \bibnamefont {England}},\
  }\href {\doibase 10.1103/PhysRevLett.116.120601} {\bibfield  {journal}
  {\bibinfo  {journal} {Phys. Rev. Lett.}\ }\textbf {\bibinfo {volume} {116}},\
  \bibinfo {pages} {120601} (\bibinfo {year} {2016})}\BibitemShut {NoStop}%
\bibitem [{\citenamefont {Proesmans}\ and\ \citenamefont {den
  Broeck}(2017)}]{Proesmans2017}%
  \BibitemOpen
  \bibfield  {author} {\bibinfo {author} {\bibfnamefont {K.}~\bibnamefont
  {Proesmans}}\ and\ \bibinfo {author} {\bibfnamefont {C.~V.}\ \bibnamefont
  {den Broeck}},\ }\href {\doibase 10.1209/0295-5075/119/20001} {\bibfield
  {journal} {\bibinfo  {journal} {EPL}\ }\textbf {\bibinfo {volume} {119}},\
  \bibinfo {pages} {20001} (\bibinfo {year} {2017})}\BibitemShut {NoStop}%
\bibitem [{\citenamefont {Hasegawa}\ and\ \citenamefont
  {Van~Vu}(2019)}]{Hasegawa2019}%
  \BibitemOpen
  \bibfield  {author} {\bibinfo {author} {\bibfnamefont {Y.}~\bibnamefont
  {Hasegawa}}\ and\ \bibinfo {author} {\bibfnamefont {T.}~\bibnamefont
  {Van~Vu}},\ }\href {\doibase 10.1103/PhysRevLett.123.110602} {\bibfield
  {journal} {\bibinfo  {journal} {Phys. Rev. Lett.}\ }\textbf {\bibinfo
  {volume} {123}},\ \bibinfo {pages} {110602} (\bibinfo {year}
  {2019})}\BibitemShut {NoStop}%
\bibitem [{\citenamefont {Agarwalla}\ and\ \citenamefont
  {Segal}(2018)}]{Agarwalla2018}%
  \BibitemOpen
  \bibfield  {author} {\bibinfo {author} {\bibfnamefont {B.~K.}\ \bibnamefont
  {Agarwalla}}\ and\ \bibinfo {author} {\bibfnamefont {D.}~\bibnamefont
  {Segal}},\ }\href {\doibase 10.1103/PhysRevB.98.155438} {\bibfield  {journal}
  {\bibinfo  {journal} {Phys. Rev. B}\ }\textbf {\bibinfo {volume} {98}},\
  \bibinfo {pages} {155438} (\bibinfo {year} {2018})}\BibitemShut {NoStop}%
\bibitem [{\citenamefont {Macieszczak}, \citenamefont {Brandner},\ and\
  \citenamefont {Garrahan}(2018)}]{Macieszczak2018}%
  \BibitemOpen
  \bibfield  {author} {\bibinfo {author} {\bibfnamefont {K.}~\bibnamefont
  {Macieszczak}}, \bibinfo {author} {\bibfnamefont {K.}~\bibnamefont
  {Brandner}}, \ and\ \bibinfo {author} {\bibfnamefont {J.~P.}\ \bibnamefont
  {Garrahan}},\ }\href {\doibase 10.1103/PhysRevLett.121.130601} {\bibfield
  {journal} {\bibinfo  {journal} {Phys. Rev. Lett.}\ }\textbf {\bibinfo
  {volume} {121}},\ \bibinfo {pages} {130601} (\bibinfo {year}
  {2018})}\BibitemShut {NoStop}%
\bibitem [{\citenamefont {Brandner}, \citenamefont {Hanazato},\ and\
  \citenamefont {Saito}(2018)}]{Brandner2018}%
  \BibitemOpen
  \bibfield  {author} {\bibinfo {author} {\bibfnamefont {K.}~\bibnamefont
  {Brandner}}, \bibinfo {author} {\bibfnamefont {T.}~\bibnamefont {Hanazato}},
  \ and\ \bibinfo {author} {\bibfnamefont {K.}~\bibnamefont {Saito}},\ }\href
  {\doibase 10.1103/PhysRevLett.120.090601} {\bibfield  {journal} {\bibinfo
  {journal} {Phys. Rev. Lett.}\ }\textbf {\bibinfo {volume} {120}},\ \bibinfo
  {pages} {090601} (\bibinfo {year} {2018})}\BibitemShut {NoStop}%
\bibitem [{\citenamefont {Miller}\ and\ \citenamefont
  {Anders}(2018)}]{Miller2018}%
  \BibitemOpen
  \bibfield  {author} {\bibinfo {author} {\bibfnamefont {H.~J.}\ \bibnamefont
  {Miller}}\ and\ \bibinfo {author} {\bibfnamefont {J.}~\bibnamefont
  {Anders}},\ }\href@noop {} {\bibfield  {journal} {\bibinfo  {journal} {Nat.
  Commun.}\ }\textbf {\bibinfo {volume} {9}},\ \bibinfo {pages} {2203}
  (\bibinfo {year} {2018})}\BibitemShut {NoStop}%
\bibitem [{\citenamefont {Kindermann}\ and\ \citenamefont
  {Pilgram}(2004)}]{Kindermann2004}%
  \BibitemOpen
  \bibfield  {author} {\bibinfo {author} {\bibfnamefont {M.}~\bibnamefont
  {Kindermann}}\ and\ \bibinfo {author} {\bibfnamefont {S.}~\bibnamefont
  {Pilgram}},\ }\href {\doibase 10.1103/PhysRevB.69.155334} {\bibfield
  {journal} {\bibinfo  {journal} {Phys. Rev. B}\ }\textbf {\bibinfo {volume}
  {69}},\ \bibinfo {pages} {155334} (\bibinfo {year} {2004})}\BibitemShut
  {NoStop}%
\bibitem [{\citenamefont {Holubec}\ and\ \citenamefont
  {Ryabov}(2017)}]{Holubec2017a}%
  \BibitemOpen
  \bibfield  {author} {\bibinfo {author} {\bibfnamefont {V.}~\bibnamefont
  {Holubec}}\ and\ \bibinfo {author} {\bibfnamefont {A.}~\bibnamefont
  {Ryabov}},\ }\href {\doibase 10.1103/PhysRevE.96.030102} {\bibfield
  {journal} {\bibinfo  {journal} {Phys. Rev. E}\ }\textbf {\bibinfo {volume}
  {96}},\ \bibinfo {pages} {030102} (\bibinfo {year} {2017})}\BibitemShut
  {NoStop}%
\bibitem [{\citenamefont {Campisi}\ and\ \citenamefont
  {Fazio}(2016)}]{Campisi2016}%
  \BibitemOpen
  \bibfield  {author} {\bibinfo {author} {\bibfnamefont {M.}~\bibnamefont
  {Campisi}}\ and\ \bibinfo {author} {\bibfnamefont {R.}~\bibnamefont
  {Fazio}},\ }\href {\doibase doi:10.1038/ncomms11895} {\bibfield  {journal}
  {\bibinfo  {journal} {Nat. Commun.}\ }\textbf {\bibinfo {volume} {7}},\
  \bibinfo {pages} {11895} (\bibinfo {year} {2016})}\BibitemShut {NoStop}%
\end{thebibliography}
%

\end{document}